\documentclass[a4paper,fleqn]{cas-dc}

\usepackage[authoryear,longnamesfirst]{natbib}

\def\tsc#1{\csdef{#1}{\textsc{\lowercase{#1}}\xspace}}
\tsc{WGM}
\tsc{QE}

\begin{document}
\let\WriteBookmarks\relax
\def\floatpagepagefraction{1}
\def\textpagefraction{.001}

\shorttitle{Luciano-Saridakis holographic dark energy}
\shortauthors{Leizerovich et al.}

\title[mode=title]{Observational constraints on Luciano-Saridakis holographic dark energy}

\tnotemark[1]
\tnotetext[1]{}

\affiliation[1]{organization={Universidad de Buenos Aires, Facultad de Ciencias Exactas y Naturales, Departamento de F\'{\i}sica},
            city={Buenos Aires},
            country={Argentina}}

\affiliation[2]{organization={CONICET -- Universidad de Buenos Aires, Instituto de F\'{\i}sica de Buenos Aires (IFIBA)},
            city={Buenos Aires},
            country={Argentina}}

\affiliation[3]{organization={Department of Chemistry, Physics and Environmental and Soil Sciences, Escola Polit\`{e}cnica Superior, Universidad de Lleida},
            addressline={Av. Jaume II, 69},
            postcode={25001},
            city={Lleida},
            country={Spain}}

\affiliation[4]{organization={Institute for Astronomy, Astrophysics, Space Applications and Remote Sensing, National Observatory of Athens},
            postcode={15236},
            city={Penteli},
            country={Greece}}

\affiliation[5]{organization={Department of Physics, National Technical University of Athens},
            addressline={Zografou Campus},
            postcode={157 73},
            city={Athens},
            country={Greece}}

\affiliation[6]{organization={Departamento de Matem\'{a}ticas, Universidad Cat\'{o}lica del Norte},
            addressline={Avda. Angamos 0610, Casilla 1280},
            city={Antofagasta},
            country={Chile}}

\affiliation[7]{organization={CAS Key Laboratory for Researches in Galaxies and Cosmology, Department of Astronomy, University of Science and Technology of China},
            city={Hefei, Anhui 230026},
            country={P.R. China}}

\author[1,2]{Mat\'{\i}as Leizerovich}[orcid=0000-0002-6438-2285]
\cormark[1]
\ead{mleize@df.uba.ar}
\credit{Methodology, Software, Data Curation, Writing, Review and Editing}

\author[1,2]{Susana J. Landau}[orcid=0000-0003-2645-9197]
\ead{slandau@df.uba.ar}
\credit{Methodology, Data Curation, Writing, Review and Editing}

\author[3]{Giuseppe Gaetano Luciano}[orcid=0000-0002-5129-848X]
\ead{giuseppegaetano.luciano@udl.cat}
\credit{Conceptualization, Formal Analysis, Writing, Review and Editing}

\author[4,5]{Andreas Papatriantafyllou}
\ead{apapatriantafyllou@mail.ntua.gr}
\credit{Writing}

\author[4,6,7]{Emmanuel N. Saridakis}[orcid=0000-0003-1500-0874]
\ead{msaridak@noa.gr}
\credit{Conceptualization, Formal Analysis, Writing, Review}

\cortext[1]{Corresponding author}

\begin{abstract}
Holographic dark energy (HDE) models provide a natural framework for linking gravitational thermodynamics to the late-time accelerated expansion of the Universe. In this work, we investigate the observational viability of an extended HDE scenario arising from a recently proposed generalized entropy. For bounded systems, this entropy exhibits a generalized holographic scaling with two independent area contributions, giving rise to a modified HDE density that encompasses both standard HDE and $\Lambda$CDM as limiting cases. Focusing on the Hubble-horizon infrared cutoff, we constrain the model using Cosmic Chronometers, the Pantheon$^+$+SH0ES Type Ia supernova compilation, DESI DR2 baryon acoustic oscillations, and compressed Planck 2018 CMB shift parameters. We find that the model provides an excellent fit to the combined dataset and admits regions of parameter space in which the Pantheon$^+$+SH0ES and CMB constraints can be simultaneously accommodated. The preferred solutions lie close to the $\Lambda$CDM regime, although non-standard entropic contributions remain compatible with current observations. We further compare the complete realization of the model, containing both independent area contributions, with its reduced single-contribution limit, finding that both provide essentially equivalent descriptions of the data, with a mild preference for the latter. Our results establish generalized entropic HDE as a viable and theoretically motivated extension of the standard cosmological scenario and provide the first observational assessment of this cosmological framework.
\end{abstract}








\maketitle


\section{Introduction}
The vast body of cosmological data gathered over the past few decades has 
definitively confirmed that the Universe transitioned from a matter-dominated 
epoch to a late-time phase of accelerated expansion. Within the framework of the 
standard cosmological paradigm, this phenomenon is typically attributed to a 
cosmological constant ($\Lambda$), which provides the most straightforward and 
mathematically economical fit to the observations. However, despite its 
empirical success, this interpretation faces profound conceptual challenges. Most 
notably, the enormous discrepancy between the observed vacuum energy density and 
the theoretical predictions of quantum field theory, coupled with the 
possibility that cosmic acceleration is a dynamical process rather than a static 
property, has motivated the search for alternative theoretical models.

In this landscape, two primary approaches have been extensively explored. The 
first preserves general relativity as the fundamental theory of gravity while 
introducing new components into the matter sector, collectively known as dark 
energy~\cite{Copeland:2006wr,Cai:2009zp,Bamba:2012cp}. The second involves 
modifying or extending the gravitational sector itself, leading to theories 
that recover general relativity in the appropriate limits while introducing 
new degrees of freedom capable of driving the late-time accelerated 
expansion~\cite{Nojiri:2010wj,Capozziello:2011et,Cai:2015emx,CANTATA:2021asi,CosmoVerseNetwork:2025alb}.

A fundamentally different approach is offered by holographic dark energy (HDE), 
based on  the holographic 
principle~\cite{tHooft:1993dmi,Susskind:1994vu,Bousso:2002ju} 
and its cosmological applications~\cite{Fischler:1998st,Bak:1999hd,Horava:2000tb}. This perspective suggests that 
the validity of effective quantum field theory at large scales is constrained by 
gravitational considerations that link ultraviolet and infrared cutoffs. 
Consequently, the vacuum energy density is determined by a characteristic 
cosmological length scale~\cite{Cohen:1998zx,Addazi:2021xuf}, giving dark energy 
a holographic origin in which its density evolves dynamically with the 
Universe~\cite{Li:2004rb,Wang:2016och}. Holographic dark energy has been studied 
extensively, revealing a rich and viable phenomenology that remains consistent 
with the Universe's thermal history and current observational constraints~\cite{Li:2004rb,Wang:2016och,Huang:2004ai,Pavon:2005yx,Wang:2005jx,
Nojiri:2005pu,Kim:2005at,Setare:2006wh,Setare:2008hm,Sheykhi:2009dz,Li:2009bn, 
Zhang:2009un,Duran:2010hi,Lu:2009iv,Micheletti:2009jy,
Aviles:2011sfa,Zhai:2011pp,Zhang:2012uu,Zhang:2015rha,Nastase:2016sji,Mukherjee:2016lor,Zhao:2017urm,Landim:2022jgr,Li:2024bwr}.

Extensions of the HDE scenario generally follow two main paths. 
One explores different infrared cutoffs, such as the future event horizon, the 
apparent horizon, the cosmic age, or curvature-based scales such as the Ricci 
scalar and the Gauss-Bonnet 
invariant~\cite{Gong:2004fq,Saridakis:2007cy,Cai:2007us,Setare:2008bb,
Gong:2009dc,
Suwa:2009gm,Jamil:2010vr,Bouhmadi-Lopez:2011qvd,Landim:2015hqa,Shekh:2021ule,
Rudra:2022qbv}. The other focuses on modifying the 
entropy-area relationship of the cosmological horizon, since drawing on 
generalized 
statistical mechanics and quantum gravity, various entropy functionals have 
been proposed, including Tsallis, Barrow, Kaniadakis, and logarithmic 
corrections, resulting in extended HDE 
models~\cite{Pourhassan:2017cba,Saridakis:2017rdo,
Nojiri:2017opc,Saridakis:2018unr,DAgostino:2019wko,Saridakis:2020zol,
Dabrowski:2020atl,Ghaffari:2022skp,Anagnostopoulos:2020ctz,
Bhattacharjee:2020ixg,Huang:2021zgj,Drepanou:2021jiv,Luciano:2022ffn,
Nojiri:2021iko,Jusufi:2021fek,Luciano:2025fox,Hernandez-Almada:2021aiw,
Hernandez-Almada:2021rjs,Luciano:2022hhy,
Luciano:2022pzg,Yarahmadi:2024oqv,Cimdiker:2025vfn}.

A shared feature of these models is that the modified entropy is typically 
introduced macroscopically through a phenomenological deformation of the 
horizon 
entropy. Recently, however, a new two-parameter generalized entropic functional 
was introduced in~\cite{Luciano:2026ufu}, based on a revised statistical 
framework and microstate counting. Unlike previous approaches, this framework 
has a clear microscopic origin, leading to a generalized holographic scaling 
for bounded systems characterized by two independent contributions. The 
resulting 
entropy recovers the standard Bekenstein-Hawking limit while exhibiting a more 
intricate structure arising from the interplay between two distinct entropic 
sectors.

Integrated into a holographic framework, this entropy leads to an extended 
HDE scenario in which the energy density comprises two 
independent holographic contributions~\cite{Luciano:2026eiy}. Consequently, the 
traditional single-scaling HDE model is promoted to a multi-sector structure in 
which different scaling behaviors can dynamically compete. This gives rise to a 
significantly richer cosmological phenomenology than that of conventional 
models, 
allowing for a variety of dynamical phases, including quintessence and phantom 
regimes, while recovering both standard HDE and the $\Lambda$CDM paradigm as 
limiting cases.

The background cosmological implications of this framework have already been 
investigated, showing that it can successfully reproduce the thermal history of 
the Universe and account for the transition from a matter-dominated epoch to 
the present phase of accelerated expansion \cite{Luciano:2026eiy}. 
Nevertheless, any 
viable cosmological scenario must ultimately be confronted with observations. 
Therefore, assessing the physical relevance of this framework requires a direct 
comparison with current cosmological data, allowing its parameter space to be 
constrained and its predictions to be tested against the observed expansion 
history of the Universe.

In this paper, we conduct a rigorous observational study of the HDE scenario 
derived from this two-parameter entropy. By utilizing a suite 
of early- and late-Universe probes, including Cosmic Chronometers, Type Ia 
supernovae from the Pantheon+ compilation (with SH0ES calibration), baryon 
acoustic oscillations from DESI DR2, and compressed Cosmic Microwave Background 
data, we perform a comprehensive statistical analysis to constrain the model's 
free parameters. This enables us to assess its phenomenological viability, 
quantify potential deviations from $\Lambda$CDM, and determine whether the 
extended parameter space can alleviate existing tensions among cosmological 
datasets.

The paper is organized as follows. In Sec.~\ref{sec:framework}, we present the 
theoretical framework of the model. In Sec.~\ref{sec:data}, we describe the 
observational methodology and datasets employed in the analysis. In 
Sec.~\ref{sec:results}, we present the results of the statistical analysis and 
discuss their cosmological implications. Finally, in 
Sec.~\ref{sec:Conclusions}, 
we summarize our conclusions and outline possible directions for future work. 
  
\section{Holographic dark energy from a new two-parameter entropy}
\label{sec:framework}

In this section we summarize the theoretical framework of the Luciano-Saridakis 
holographic dark energy scenario. We first review the generalized two-parameter 
entropy and then present the corresponding holographic dark energy construction 
that will be tested against cosmological observations in the following sections.
  
\subsection{Generalized two-parameter entropic functional}
In the standard formulation of statistical mechanics, equilibrium systems are
described by the Boltzmann-Gibbs-Shannon (BGS) entropy,
\begin{equation}
\label{SBG}
S_{\text{BGS}}=-\kappa\sum_{i=1}^{W}p_i\ln p_i\,,
\end{equation}
where $p_i$ denotes the probability of the $i$-th microstate among the $W$
accessible configurations, and $\kappa$ sets the entropy scale (in gravitational
applications  it is typically identified with the Boltzmann 
constant)~\cite{goldstein2020gibbs}.

Although Eq.~\eqref{SBG} successfully describes a wide class of physical 
systems,
it is known to be insufficient in situations involving long-range interactions,
strong correlations, or non-Markovian dynamics, where the effective counting of
microstates departs from the standard extensive behavior.

A natural generalization of the BGS entropy can be constructed within the 
functional
framework~\cite{hanel2011comprehensive},
\begin{equation}
\label{Sf}
S_f[p]=\sum_i f(p_i)\,,
\end{equation}
for an appropriate choice of the function $f$. Imposing the four
Khinchin axioms - continuity, maximality, expandability, and separability - 
leads
uniquely to the BGS entropy~\cite{shannon1948claude,Khinchin1957}. However, the
assumption of separability is not expected to hold in systems exhibiting
non-trivial correlations or effective non-additivity. Relaxing this requirement
allows for the construction of generalized entropy functionals that remain
thermodynamically consistent while capturing a broader class of microscopic
behaviors.

In gravitational systems, and particularly in cosmological settings, entropy
plays a special role due to its holographic scaling. In the microcanonical
ensemble, the Bekenstein-Hawking entropy obeys the area law
\begin{equation}
\label{arealaw}
S_{\text{BGS}}\propto \log W \propto L^2\,,
\end{equation}
where $L$ denotes the characteristic size of the 
system~\cite{tHooft:1993dmi,Susskind:1994vu}. This scaling implies that the 
number of accessible microstates grows as
\begin{equation}
\label{W}
W=g(L)\,\xi^{L^2}\,, \qquad \xi>1\,,
\end{equation}
where $g(L)$ represents subleading contributions at large scales. In a
four-dimensional spacetime, the area-law scaling of entropy contrasts with the
volume scaling expected for an extensive thermodynamic quantity. This suggests
the presence of strong correlations among the underlying degrees of freedom and
motivates the consideration of more general entropic frameworks.

Building upon this perspective, a generalized entropy functional characterized
by two independent exponents was recently introduced in~\cite{Luciano:2026ufu}.
Within this framework, one considers
\begin{eqnarray}
\nonumber
S_{\delta,\epsilon}&=&\eta_\delta \sum_i
p_i\left(\log\frac{1}{p_i}\right)^\delta
+\eta_\epsilon \sum_i p_i\left(\log\frac{1}{p_i}\right)^\epsilon\\[1mm]
&=&\eta_\delta\left(\log W\right)^\delta+\eta_\epsilon\left(\log
W\right)^\epsilon\,,
\label{genentropyexpr}
\end{eqnarray}
for equiprobable distributions, where $\delta,\epsilon>0$. This expression
corresponds to the generalized microstate counting \cite{Luciano:2026ufu}
\begin{equation}
\label{microscaling}
W_{\delta,\epsilon}=g(L)\,\xi^{L^{2\delta}}\,\tilde{\xi}^{L^{2\epsilon}}\,,
\qquad
\xi,\tilde{\xi}>1\,,
\end{equation}
which extends the standard holographic scaling through the coexistence of two
independent contributions.

For systems bounded by a surface of area $A\sim L^2$, the above scaling yields,
in the large-$L$ limit, the generalized entropy
\begin{equation}
\label{Sde}
S_{\delta,\epsilon}=\gamma_\delta A^\delta+\gamma_\epsilon A^\epsilon\,,
\end{equation}
where $\gamma_\delta$ and $\gamma_\epsilon$ are positive constants with
dimensions $[L^{-2\delta}]$ and $[L^{-2\epsilon}]$, respectively. Hence, within 
this
framework, the exponents $(\delta,\epsilon)$ characterize departures from the
standard holographic scaling through two independent entropic sectors.

It is important to emphasize that the entropy \eqref{Sde} is not introduced at
the macroscopic level as an \emph{ad hoc} modification of the area law. Rather,
it emerges naturally from the underlying statistical description encoded in
\eqref{genentropyexpr} and the associated microstate scaling
\eqref{microscaling}, which result from a controlled relaxation of the
separability axiom. In this way, the framework establishes a direct connection
between microscopic statistical properties and effective gravitational entropy.

More generally, the scaling \eqref{microscaling} can be interpreted as the
simplest realization of a multi-scaling structure, analogous to those appearing
in complex systems with hierarchical correlations or multifractal
behavior~(see, e.g.,~\cite{hanel2011comprehensive}). The presence of two
independent exponents enriches the statistical description beyond conventional
single-parameter generalizations, as the growth of accessible microstates is no
longer governed by a unique scaling law. Instead, two distinct entropic sectors
contribute simultaneously to the counting of microscopic configurations,
allowing different scaling behaviors to coexist and compete across scales. As a
result, the entropy acquires a greater degree of flexibility while retaining a
well-defined microscopic origin, providing a natural framework for describing
systems whose effective organization cannot be captured by a single scaling
exponent.

As expected, the standard Bekenstein-Hawking entropy is consistently recovered
in the limiting cases
\begin{eqnarray}
\label{limits}
 &&\delta=1,\ \gamma_\delta=\frac{1}{4\ell_p^2},\ \gamma_\epsilon=0;\notag\\
&&\epsilon=1,\ \gamma_\delta=0,\ \gamma_\epsilon=\frac{1}{4\ell_p^2};\notag\\
&&\delta=\epsilon=1,\ \gamma_\delta=\gamma_\epsilon=\frac{1}{8\ell_p^2},
\end{eqnarray}
where $\ell_p$ is the Planck length. Furthermore,  in the special case 
$\delta=\epsilon$, the entropy \eqref{Sde} reduces
to the corresponding single-exponent power-law form, thereby recovering
well-known frameworks such as the Tsallis and Barrow 
entropies~\cite{Tsallis:2013,Barrow:2020tzx}.

\subsection{Luciano-Saridakis  holographic dark energy}
\label{HDEl}

We now proceed to construct the HDE scenario arising from
the two-parameter entropy introduced above. Throughout this work, we assume a
spatially flat, homogeneous, and isotropic Friedmann-Robertson-Walker
background geometry,
\begin{equation}
\label{FRc}
ds^{2}=-dt^{2}+a^{2}(t)\delta_{ij}dx^{i}dx^{j}\,,
\end{equation}
where $a(t)$ is the scale factor. The cosmological dynamics are then governed
by the first Friedmann equation,
\begin{equation}
\label{Fr1bFRW}
3M_p^2 H^2 = \rho_m + \rho_{DE},
\end{equation}
where $M_p$ is the reduced Planck mass, $\rho_m$ and $\rho_{DE}$ denote the
matter and dark energy densities, respectively, and
$H\equiv \dot a/a$ is the Hubble parameter, with an overdot denoting
differentiation with respect to cosmic time $t$.

In the present framework the standard Bekenstein-Hawking entropy is replaced by 
the generalized entropy~\eqref{Sde}. Hence, applying the  holographic 
prescription, one is
led to the modified dark energy density\footnote{With a slight abuse of 
notation, we denote by the same symbols
$\gamma_\delta$ and $\gamma_\epsilon$ the effective coefficients appearing in
the dark energy density, absorbing the numerical factors generated by the
holographic prescription into their definition.}~\cite{Luciano:2026eiy}
\begin{equation}
\label{GenHDE}
\rho_{DE} 
=\gamma_\delta L^{2(\delta-2)}+\gamma_\epsilon L^{2(\epsilon-2)}.
\end{equation}

Consistently, whenever the entropy \eqref{Sde} reduces to its standard form,
the expression \eqref{GenHDE} recovers the standard HDE
scenario~\cite{Li:2004rb,Wang:2016och}. In contrast, for generic values of the
two entropic exponents, one obtains a genuine two-sector extension of
HDE. 
Furthermore, if $\epsilon=2$ with $\gamma_\delta=0$, or
$\delta=2$ with $\gamma_\epsilon=0$, the dark energy density becomes
constant and the model reduces to the $\Lambda$CDM scenario. More generally,
for $\delta=\epsilon=2$ one obtains
$\rho_{DE}=\gamma_\delta+\gamma_\epsilon=\text{const.}$,
which is likewise equivalent to a cosmological constant.

Regarding the choice of the infrared cutoff $L$, several possibilities have
been explored in the HDE literature. In the present work, we restrict the 
analysis to the Hubble-horizon realization, namely $L=H^{-1}$. This choice is
determined entirely by the local background expansion and provides the simplest
implementation of the holographic prescription, allowing the resulting dark
energy density to be written directly in terms of the Hubble rate.

On the other hand, the future event horizon 
represents another physically motivated infrared cutoff, as it encodes the
global future causal structure of spacetime and is known to support accelerated
expansion in standard HDE. However, in the present two-parameter entropic
framework, this choice leads to a considerably more involved dynamical system,
whose fully general analytical treatment is not straightforward 
\cite{Luciano:2026eiy}. A detailed
observational investigation of the future-event-horizon realization is therefore
left for future work.

We proceed by considering the choice $L=H^{-1}$. Substituting this 
into Eq.
\eqref{GenHDE}, the dark energy density becomes
\begin{equation}
\label{GenHDEHubble}
\rho_{DE}=
\gamma_\delta H^{2(2-\delta)}+\gamma_\epsilon H^{2(2-\epsilon)}.
\end{equation}
Assuming that the matter sector is separately conserved, the dark energy sector
satisfies an independent conservation equation. Using 
$\dot{\rho}_{DE}+3H\rho_{DE}(1+w_{DE})=0$, one obtains the following expression 
for the dark-energy equation-of-state
parameter:
\begin{equation}
\label{wderelation}
w_{DE}=-1+\frac{2\dot{H}\left[\gamma_\epsilon(\epsilon-2)H^{2\delta}
+\gamma_\delta(\delta-2)H^{2\epsilon}\right]}
{3H^2\left(\gamma_\epsilon H^{2\delta}+\gamma_\delta H^{2\epsilon}\right)}.
\end{equation}
Therefore, the generalized entropic structure naturally gives rise to a
dynamical effective equation of state, with its evolution determined by the
interplay between the two underlying scaling sectors.

As noted above, certain parameter configurations are of particular interest. In
particular, for $\epsilon=2$ one finds
\begin{equation}
\label{GenHDEHubblespec1}
\rho_{DE}=
\gamma_\delta H^{2(2-\delta)}+\gamma_\epsilon\,,
\end{equation}
corresponding to a dynamical correction to the $\Lambda$CDM dark energy sector.
In the further limit $\gamma_\delta=0$, the dark energy density becomes
strictly constant and the $\Lambda$CDM scenario is recovered exactly.

Similarly, for $\delta=2$ one obtains
\begin{equation}
\label{GenHDEHubblespec2}
\rho_{DE}=
\gamma_\epsilon H^{2(2-\epsilon)}+\gamma_\delta\,,
\end{equation}
representing a complementary deformation of the $\Lambda$CDM dark energy
sector. The standard $\Lambda$CDM scenario is recovered in the limit
$\gamma_\epsilon=0$.

In order to perform the cosmological analysis, it is convenient to introduce
the standard density parameters
\begin{equation}
\Omega_m\equiv\frac{\rho_m}{3M_p^2H^2}\,,\qquad
\Omega_{DE}\equiv\frac{\rho_{DE}}{3M_p^2H^2}\,.
\label{ODE}
\end{equation}
Furthermore, we reparametrize Eq. \eqref{GenHDEHubble} in terms of dimensionless 
amplitudes:
\begin{equation}
  \frac{\rho_{DE}}{H_0^2}
  = \alpha_\delta\, E^{n_\delta} 
  + \alpha_\epsilon\, E^{n_\epsilon}\,,
  \label{eq:rhoDE_dimless}
\end{equation}
where $E \equiv H/H_0$, $\alpha_\delta= \gamma_\delta\,H_0^{2(1-\delta)} $,  
$\alpha_\epsilon= \gamma_\epsilon\,H_0^{2(1-\epsilon)} $, and the exponents are
\begin{equation}
  n_\delta = 2(2-\delta)\,,\qquad n_\epsilon = 2(2-\epsilon)\,.
  \label{eq:exponents}
\end{equation}
At $z=0$ ($E=1$), one finds
\begin{equation}
\Omega_{{DE},0}
=
\frac{\alpha_\delta+\alpha_\epsilon}{3M_p^2}\,.
\label{eq:budget}
\end{equation} 
As can be seen from
Eq.~\eqref{eq:rhoDE_dimless}, strong degeneracies arise between
$\alpha_\delta$ and $\alpha_\epsilon$, as well as between $\delta$ and
$\epsilon$. These degeneracies significantly impair the efficiency and
convergence of the statistical inference procedure we perform in the following. 
To mitigate this issue, we
restrict the parameter space by fixing $\delta=2$ throughout the observational
analysis. In addition, we consider the special case $\alpha_\delta=0$, which
corresponds to the one-sector limit of the model.

 \section{Observational framework and data}\label{sec:data}

In this section, we present the observational probes used to constrain the
Luciano-Saridakis HDE scenario. Our analysis incorporates Type Ia supernovae
(SNIa), Cosmic Chronometers (CC), Baryon Acoustic Oscillations (BAO), and
compressed Cosmic Microwave Background (CMB) information through shift
parameters. These observables probe the background expansion history of the
Universe and depend primarily on the Hubble parameter $H(z)$ and the comoving
distance,
\begin{equation}
\chi(z)=c\int_{0}^{z}\frac{dz'}{H(z')}\,,
\end{equation}
where the speed of light $c$ has been restored explicitly. Since we consider a 
spatially flat universe, the comoving distance coincides
with the transverse comoving distance, namely $D_M(z)=\chi(z)$.

The CC method~\cite{Moresco:2012by,Moresco:2015cya,Moresco:2016mzx}
provides direct measurements of the Hubble parameter $H(z)$ through
differential age estimates of passively evolving galaxies. In this work, we
use the compilation reported in Table IX of Ref.~\cite{Chantada2023}, taking
into account the full covariance matrix, including the non-diagonal
correlations discussed in Ref.~\cite{2020ApJ...898...82M}.

SNIa constitute one of the most powerful distance probes in
cosmology, as their homogeneous spectral properties and standardizable light
curves enable precise determinations of the luminosity distance. In this work,
we employ the full Pantheon$^+$ compilation, comprising 1701 supernovae in the
redshift range $0.001<z<2.3$. The dataset also incorporates the
Cepheid-calibrated SH0ES information at very low redshift ($z<0.01$). We
therefore refer to it as PPS (Pantheon$^+$ + 
SH0ES)~\cite{Scolnic:2021amr,Brout:2022vxf}. The relevant observable is the 
distance modulus,
\begin{equation}
\mu(z)=5\log_{10}\left(\frac{D_L(z)}{\mathrm{Mpc}}\right)+25,
\end{equation}
where $D_L(z)$ is the luminosity distance, given by
\begin{equation}
D_L(z)=(1+z)D_M(z).
\end{equation}

On the other hand, BAO originate from sound waves propagating in
the primordial photon-baryon plasma and imprint a characteristic scale on the
late-time distribution of matter. This scale is set by the sound horizon at the
drag epoch, $r_d$, and serves as a standard ruler for cosmological distance
measurements. In this work, we employ the recent DESI DR2 BAO 
dataset~\cite{DESI:2025zgx},
which provides constraints on both $D_H(z)/r_d$ and $D_M(z)/r_d$, where the
Hubble distance is defined as
\begin{equation}
D_H(z)=\frac{c}{H(z)}\,.
\end{equation}
The drag-epoch sound horizon is defined as $r_d=r_s(z_d)$, with
\begin{equation}
r_s(z)=\int_{z}^{\infty}\frac{c_s(z')}{H(z')}\,dz'\,,
\end{equation}
where $c_s(z)$ is the sound speed in the photon-baryon fluid.

For the CMB, rather than using the full
temperature and polarization anisotropy spectra, we employ the CMB shift
parameters, which provide a compressed representation of the geometrical
information encoded in the background expansion history. These are given by
\begin{equation}
R=\sqrt{\Omega_{m0}}\,\frac{H_0}{c}\,D_M(z^*)\,,
\qquad
l_A=\pi\,\frac{D_M(z^*)}{r_s(z^*)}\,,
\end{equation}
where $z^*$ denotes the redshift of recombination.

While $R$ probes the distance to the last-scattering surface, $l_A$ encodes the
angular scale of the sound horizon at recombination. In practice, these
quantities, together with the baryon density parameter
$\omega_b=\Omega_b h^2$, capture most of the CMB information relevant to
background cosmology.  In our analysis, we adopt the values reported in
Ref.~\cite{2024PhLB..85438717L}, derived from the Planck 2018 TT, TE, EE +
lowE likelihood~\cite{Planck:2019nip}.

\begin{figure}[!]
\centering
\includegraphics[width=0.54\textwidth]{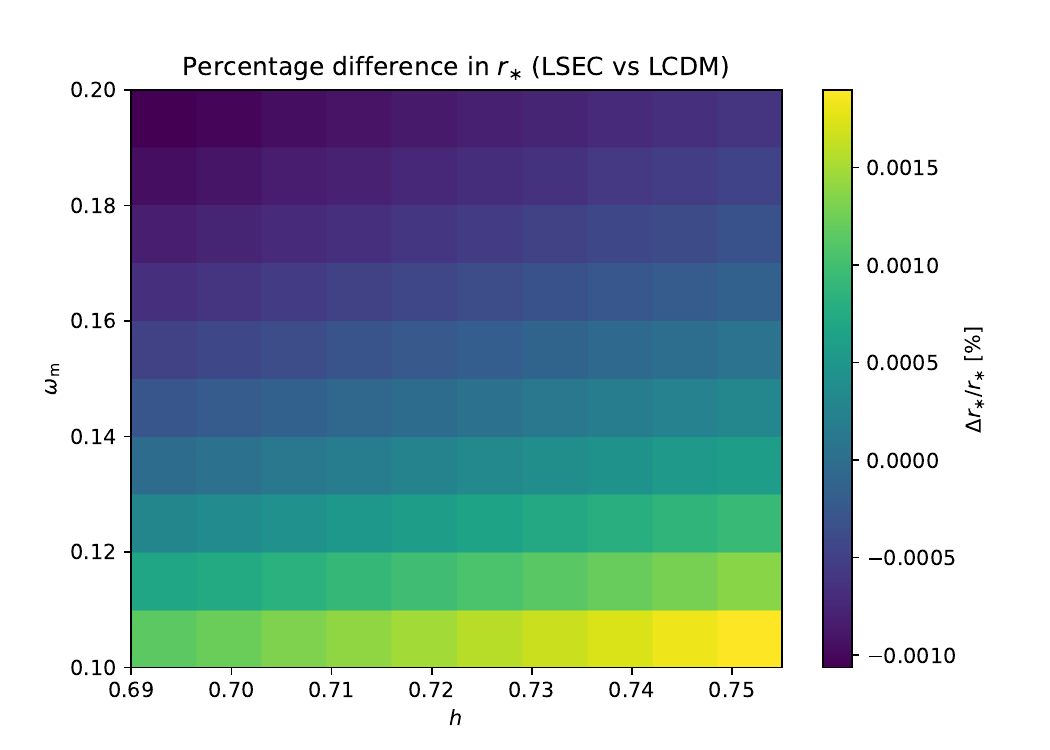}
\caption{
{\it{Relative difference in the sound horizon at the drag epoch, $r_s$,
between the HDE bestfit model and the $\Lambda$CDM reference analysis. }}}
\label{fig:rs_difference_hde_vs_lcdm}
\end{figure}

\begin{figure}[!]
\centering
\includegraphics[width=0.54\textwidth]{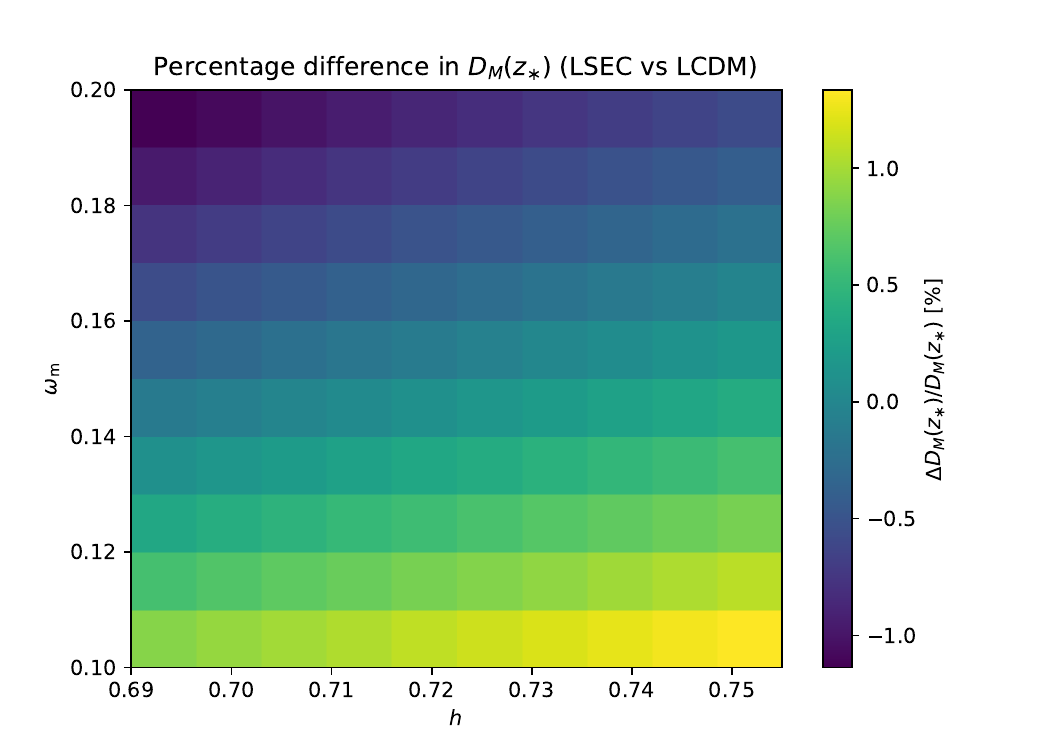}
\caption{{\it{Relative difference in the transverse comoving distance to
recombination, $D_M(z*)$, between the HDE bestfit model and the $\Lambda$CDM reference
analysis.}}}
\label{fig:DM_difference_hde_vs_lcdm}
\end{figure}

We note, however, that these compressed observables are obtained under the
assumption that the underlying cosmological model  is
$\Lambda$CDM.  To this end, we have
checked that key quantities entering the CMB observables, such as the sound
horizon and the comoving distance to the last-scattering surface, differ only
negligibly from their $\Lambda$CDM counterparts. This is explicitly illustrated
in Figs.~\ref{fig:rs_difference_hde_vs_lcdm} and
\ref{fig:DM_difference_hde_vs_lcdm}.

\begin{figure*}[t]
\centering
\includegraphics[width=0.8\textwidth]{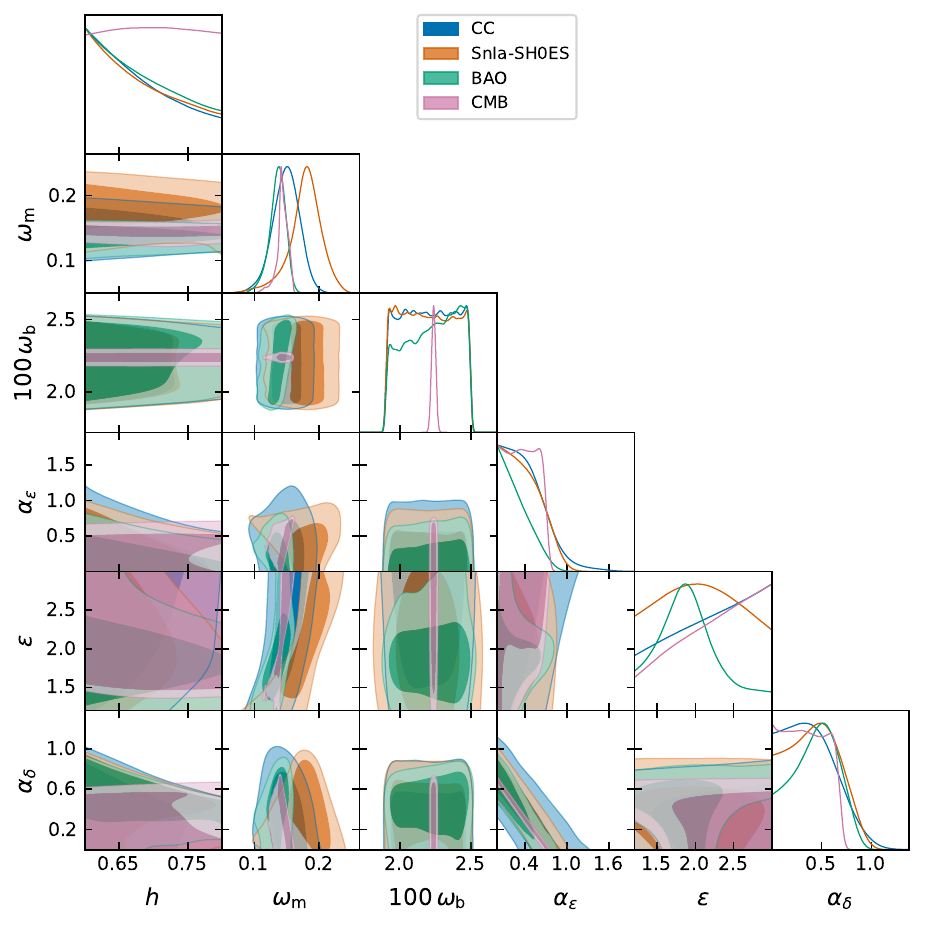}
\caption{{\it{Posterior distributions of the HDE model parameters (for
$\alpha_\delta=0$) obtained from independent analyses of the Planck 2018 CMB
shift parameters, Cosmic Chronometers (CC), Pantheon$^+$ + SH0ES (PPS), and
DESI DR2 BAO datasets. The diagonal panels display the marginalized
one-dimensional posterior distributions, while the off-diagonal panels show the
corresponding two-dimensional confidence contours. The inner and outer contours
denote the 68\% and 95\% confidence regions, respectively.}}}
\label{fig:triangle_individual}
\end{figure*}

Finally, the sampled parameters and their corresponding flat priors are
\begin{equation}
\begin{array}{lll}
h \in [0.6,\,0.8]\,,\,\, &
\omega_m \in [0.05,\,0.6]\,,\,\, &
\omega_b \in [0.019,\,0.025]\,, \\[1mm]
\epsilon \in [1.2,\,3.0]\,,\,\, &
\alpha_\epsilon \in [0.0,\,5.0]\,,\,\, &
\alpha_\delta \in [0.0,\,5.0]\,.
\end{array}
\label{eq:priors}
\end{equation}
The amplitudes $\alpha_\epsilon$ and $\alpha_\delta$ are quoted in the
normalization of Eq.~\eqref{eq:rhoDE_dimless}. In this convention, the factor
setting their dimensions is absorbed into the definition of the effective
coefficients, so that they are sampled as dimensionless parameters in the
numerical analysis.

For comparison, the $\Lambda$CDM reference analysis was performed using the same
datasets and cosmological priors whenever applicable.

\section{Observational constraints}\label{sec:results}

\begin{figure*}[t]
\centering
\includegraphics[width=0.6\textwidth]{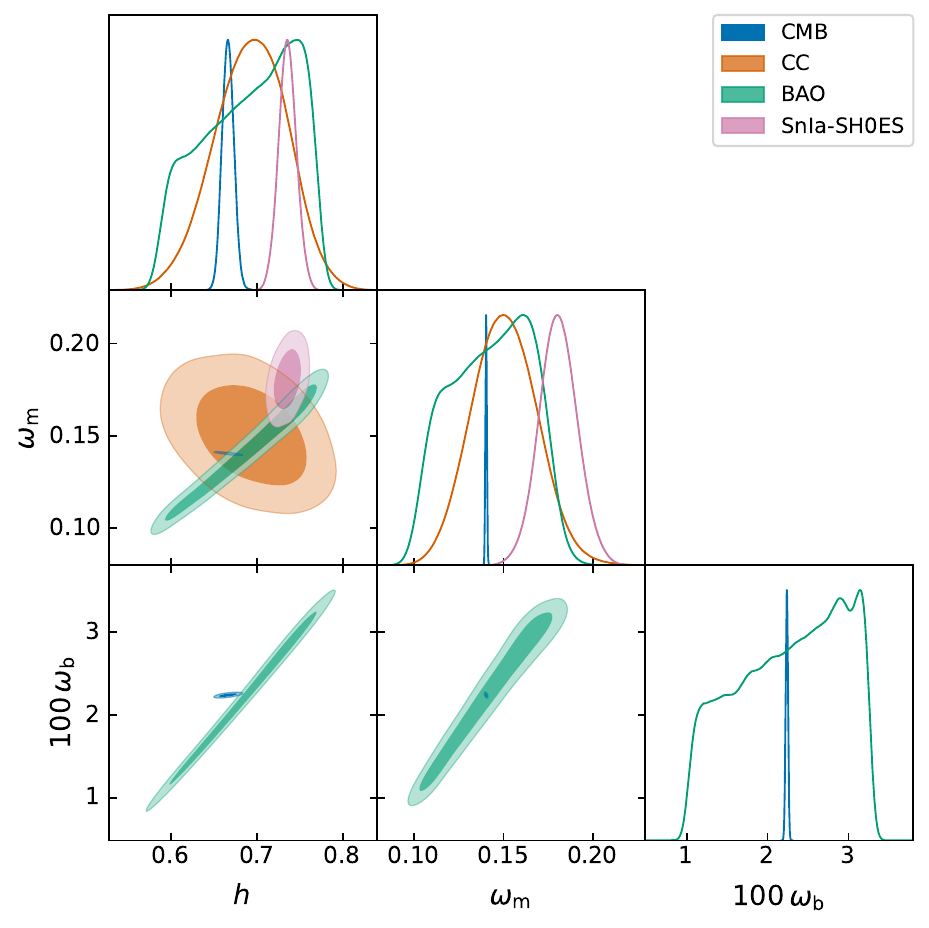}
\caption{{\it{Posterior distributions of the $\Lambda$CDM model parameters 
obtained
separately from the Planck 2018 CMB shift parameters, Cosmic Chronometers (CC),
Pantheon$^+$ + SH0ES (PPS), and DESI DR2 BAO datasets. The diagonal panels show
the marginalized one-dimensional posterior distributions, while the
off-diagonal panels present the corresponding two-dimensional confidence
contours. The inner and outer contours correspond to the 68\% and 95\%
confidence regions, respectively.}}}
\label{fig:triangle_lcdm}
\end{figure*}

We perform a Markov Chain Monte Carlo (MCMC) analysis using the
\texttt{COBAYA} framework~\cite{Torrado:2020dgo,2019ascl.soft10019T}, in
combination with a suitably modified version of the \texttt{CLASS} Boltzmann
solver~\cite{2011JCAP...07..034B}. The explored parameter space consists of the
 parameters
$(h,\omega_m,\omega_b,\alpha_\epsilon,\epsilon,\alpha_\delta)$, sampled within
the prior ranges specified above.
In all runs, we fix the present CMB temperature to
$T_{\rm CMB}=2.7255\,\mathrm{K}$ and assume a standard relativistic sector with
massless neutrinos and an effective number of relativistic species
$N_{\rm eff}=3.046$. Chain convergence is monitored using the Gelman-Rubin
criterion, requiring $R-1<10^{-2}$, while the first $30\%$ of samples in each
chain are discarded as burn-in. 
 
 \subsection{Individual dataset analyses}

 Before performing the combined analysis, we first examine the constraints 
obtained from each observational probe separately.  As shown in
Fig.~\ref{fig:triangle_individual}, the resulting confidence regions are
mutually consistent across the different probes. The figure also shows that the
CMB shift parameters provide the most stringent constraints and play a crucial 
role in
breaking the parameter degeneracies present in the other datasets.

We also carried out the analysis using each data set separately, assuming
$\Lambda$CDM as the underlying cosmological model. As anticipated from previous
studies, the CMB and PPS datasets are not consistent within this framework, 
as   can be seen in 
Fig.~\ref{fig:triangle_lcdm}. Consequently, a joint analysis of all datasets
is not feasible in the context of $\Lambda$CDM. Conversely, the HDE analysis
reveals the existence of regions in parameter space where the CMB shift
parameters and PPS constraints become mutually consistent. This does not
contradict previous results, since those regions are not favored by the data
when $\Lambda$CDM is assumed to be the underlying cosmological model.

\subsection{Joint analysis }
\label{sec:joint_hde}

\begin{figure*}[t]
\centering
\includegraphics[width=0.78\textwidth]{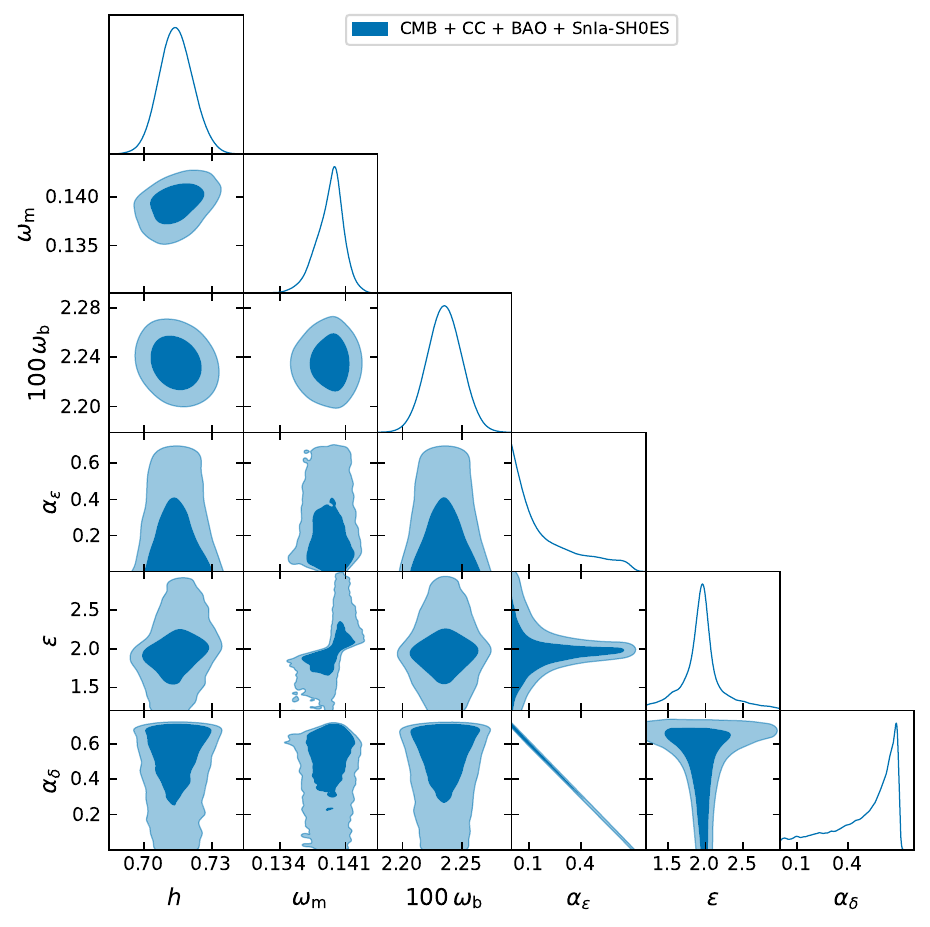}
\caption{{\it{Posterior distributions of the HDE model parameters obtained from 
the
joint analysis of the Planck 2018 CMB shift parameters, Cosmic Chronometers
(CC), Pantheon$^+$ + SH0ES (PPS), and DESI DR2 BAO datasets. The diagonal panels
show the marginalized one-dimensional posterior distributions, while the
off-diagonal panels present the corresponding two-dimensional confidence
contours. The inner and outer contours correspond to the 68\% and 95\%
confidence regions, respectively.}}}
\label{fig:triangle_joint}
\end{figure*}

The joint analysis combining all datasets (CC $+$ PPS $+$ DESI DR2 BAO
$+$ CMB shift parameters) yields the constraints summarized in
Table~\ref{tab:joint} and Fig.~\ref{fig:triangle_joint}. In this case, we
fix $\delta=2$ in order to avoid degeneracies with $\epsilon$.

\begin{table}[ht]
\centering
\caption{Joint HDE constraints from CC $+$ PPS $+$ DESI~DR2 BAO $+$ CMB shift 
parameters.}
\label{tab:joint}
\begin{tabular}{l c c}
\toprule
Parameter & 68\% limits & 95\% limits \\
\midrule

$h$                 & $0.7142\pm 0.0075$          & $0.714^{+0.015}_{-0.014}$   
 \\
$\omega_m$          & $0.1393^{+0.0016}_{-0.0012}$          & 
$0.1393^{+0.0027}_{-0.0034}$ \\
$100\,\omega_b$     & $2.235\pm 0.015$            & $2.235^{+0.029}_{-0.028}$   
\\
$\epsilon$          & $1.96^{+0.21}_{-0.23}$            & 
$1.96^{+0.74}_{-0.59}$ 
  
 \\
$\alpha_\epsilon$   & $< 0.250$          & 
$< 0.596$ \\
$\alpha_\delta$   & $0.500^{+0.21}_{-0.056}$          & 
$0.50^{+0.21}_{-0.41}$ \\
\bottomrule
\end{tabular}
\end{table}

The baryon density parameter $\omega_b$ is fully consistent with the Planck
$\Lambda$CDM determination at the $1\sigma$ level, while $\omega_m$ remains
compatible within $2\sigma$. Overall, these results indicate broad agreement
with the standard cosmological model, while leaving room for non-trivial
deviations associated with the generalized holographic dark energy sector.

Although the model favors values of $\epsilon \sim 2$, i.e. close to the
$\Lambda$CDM limit, departures from the standard cosmological constant scenario
are not completely excluded by current observational constraints.
In this respect, values of $\alpha_\delta \sim 0.7$ combined with
$\alpha_\epsilon \approx 0$ reproduce the cosmological constant behavior. The
data favor small values of $\alpha_\epsilon$, compatible with zero within the
uncertainties (see Table~\ref{tab:joint}), while constraining
$\alpha_\delta$ to $0.500^{+0.21}_{-0.056}$ at the 68\% confidence level.
Overall, the preferred region of parameter space lies close to the
$\Lambda$CDM limit, although a residual contribution from the generalized
entropic sector remains observationally viable.

Furthermore, the inferred Hubble parameter is
$H_0 = 71.42 \pm 0.75~\mathrm{km\,s^{-1}\,Mpc^{-1}}$. This value lies between
the Planck $\Lambda$CDM determination ($H_0 \approx 
67~\mathrm{km\,s^{-1}\,Mpc^{-1}}$)
and the SH0ES measurement, reflecting the impact of the SH0ES calibration
included in the Pantheon$+$ dataset.

\begin{figure*}[t]
\centering
\includegraphics[width=0.75\textwidth]{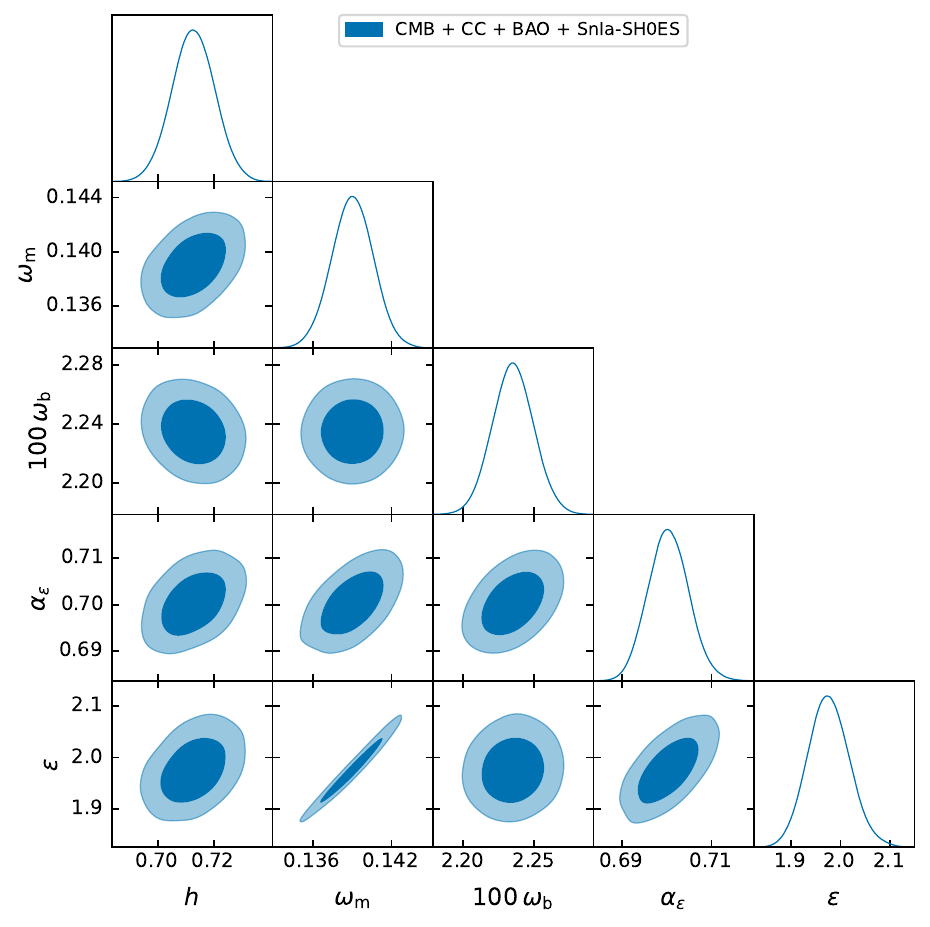}
\caption{{\it{Posterior distributions of the HDE model parameters 
obtained from the
joint analysis of the Planck 2018 CMB shift parameters, Cosmic Chronometers
(CC), Pantheon$^+$ + SH0ES (PPS), and DESI DR2 BAO datasets, assuming
$\alpha_\delta=0$. The diagonal panels show the marginalized one-dimensional
posterior distributions, while the off-diagonal panels present the
corresponding two-dimensional confidence contours. The inner and outer contours
correspond to the 68\% and 95\% confidence regions, respectively.}}}
\label{fig:triangle_joint_alpha_delta_0}
\end{figure*}

We further investigate the restricted scenario obtained by setting
$\alpha_\delta=0$ in Eq.~\eqref{eq:rhoDE_dimless}. The joint constraints derived
from the combination of CC, Pantheon$+$SH0ES, DESI DR2 BAO, and CMB shift
parameters are summarized in
Table~\ref{tab:joint_analysis_alpha_delta_0} and
Fig.~\ref{fig:triangle_joint_alpha_delta_0}.

\begin{table}[ht]
\centering
\caption{Constraints on the model parameters with $\alpha_\delta=0$.}
\label{tab:joint_analysis_alpha_delta_0}
\begin{tabular}{l c c}
\toprule
Parameter & 68\% limits & 95\% limits \\
\midrule

{\boldmath$h$}
& $0.7129\pm 0.0080$
& $0.713^{+0.016}_{-0.016}$ \\

{\boldmath$\omega_\mathrm{m}$}
& $0.1391\pm 0.0017$
& $0.1391^{+0.0034}_{-0.0032}$ \\

{\boldmath$\omega_\mathrm{b}$}
& $0.02235\pm 0.00015$
& $0.02235^{+0.00029}_{-0.00029}$ \\

{\boldmath$\epsilon$}
& $1.979^{+0.039}_{-0.048}$
& $1.979^{+0.089}_{-0.087}$ \\

{\boldmath$\alpha_\epsilon$}
& $0.7006\pm 0.0048$
& $0.7006^{+0.0097}_{-0.0088}$ \\

$\omega_\mathrm{c}$
& $0.1168\pm 0.0017$
& $0.1168^{+0.0034}_{-0.0032}$ \\

\bottomrule
\end{tabular}

\end{table}

As we observe, the standard cosmological parameters are tightly constrained and 
remain fully
consistent with the values obtained in the extended HDE scenario. This
indicates that the two models provide a comparably good description of the
current cosmological observations. 
Additionally, we find 
$\epsilon = 1.979^{+0.039}_{-0.048}$, while
$\alpha_\epsilon = 0.7006 \pm 0.0048$ at the 68\% confidence level.
The relatively tight bounds on these parameters demonstrate the constraining
power of the combined dataset and significantly restrict the viable parameter
space of the model.

To assess the role of the additional parameter $\alpha_\delta$, we compare the
extended HDE model with the restricted case $\alpha_\delta=0$ using goodness-of-fit
and information criteria. The minimum chi-square values are nearly identical,
with $\chi^2_{\rm min}({\rm HDE}) = 1500.65$ and
$\chi^2_{\rm min}(\alpha_\delta=0)=1500.63$, corresponding to a negligible
difference $\Delta\chi^2 \simeq 0.02$. This indicates that the extended and
restricted realizations provide an essentially equivalent description of the
current cosmological observations.

This conclusion is further supported by the Akaike \cite{Akaike:1974vps} and 
Bayesian \cite{Schwarz:1978tpv}
information
criteria. Defining
\begin{equation}
\Delta {\rm AIC} = {\rm AIC}_{\alpha_\delta\neq0}
- {\rm AIC}_{\alpha_\delta=0},
\end{equation}
and
\begin{equation}
\Delta {\rm BIC} = {\rm BIC}_{\alpha_\delta\neq0}
- {\rm BIC}_{\alpha_\delta=0},
\end{equation}
we obtain $\Delta {\rm AIC}=2.02$ and $\Delta {\rm BIC}=7.51$. 
These values indicate that the current data do not provide compelling  
statistical support for the additional parameter $\alpha_\delta$ within the HDE 
framework. 
In particular, while the AIC suggests only a mild preference for the reduced 
model,  the BIC favors it more strongly. Nevertheless, the extended scenario 
remains fully compatible with the observations.

In summary, the above analysis shows that present cosmological observations are
consistent with both the single-sector and the two-sector realizations of the
entropic framework.

\section{Conclusions}\label{sec:Conclusions}

In this work, we have performed a comprehensive observational analysis of the
Luciano-Saridakis holographic dark energy (HDE) scenario, which is based on a
recently proposed two-parameter generalized entropic functional with a well-defined
microscopic origin. By combining four independent cosmological probes - Cosmic
Chronometers, Type Ia supernovae from the Pantheon$^+$+SH0ES compilation,
baryon acoustic oscillations from DESI DR2, and compressed CMB shift
parameters derived from Planck 2018 - we have derived stringent constraints on
the model parameters and assessed the observational viability of this extended
holographic framework.

The joint analysis of the full HDE model, with $\delta=2$ fixed in order to
avoid parameter degeneracies, yields a Hubble constant
$H_0 = 71.42 \pm 0.75~{\rm km\,s^{-1}\,Mpc^{-1}}$. This value lies between the
Planck $\Lambda$CDM determination and the SH0ES local measurement, reflecting
the impact of the SH0ES Cepheid calibration included in the Pantheon$^+$
dataset. The standard cosmological parameters $\omega_m$ and $\omega_b$ remain
consistent with the Planck $\Lambda$CDM values within $2\sigma$ and $1\sigma$,
respectively, indicating that the generalized entropic extension preserves the
successful description of the matter sector provided by the standard
cosmological model.

The entropic exponent $\epsilon = 1.96^{+0.21}_{-0.23}$ at the 68\% confidence
level is compatible with the $\Lambda$CDM limiting value $\epsilon = 2$.
Although the preferred region of parameter space lies close to the standard
cosmological constant scenario, the current constraints still allow for modest
departures from it. The amplitude $\alpha_\epsilon$ is consistent with zero,
while $\alpha_\delta = 0.500^{+0.21}_{-0.056}$ accounts for the dominant
contribution to the dark energy sector. Overall, the results indicate that the
generalized entropic framework provides a cosmologically viable extension of
standard HDE, with the $\Lambda$CDM behavior emerging as a particular limit
within a broader entropic parameter space.

A key result of our analysis is the comparison between the full HDE model and
its restricted subcase with $\alpha_\delta = 0$. Both realizations yield
virtually identical minimum chi-square values ($\Delta\chi^2 \simeq 0.02$),
demonstrating that the extended entropic framework remains fully compatible
with current cosmological observations. The information criteria give
$\Delta\text{AIC}=2.02$ and $\Delta\text{BIC}=7.51$, indicating that the
present data do not provide a statistically significant preference for the
additional parameter $\alpha_\delta$, at least within the HDE framework. Nevertheless, the extended scenario
achieves a fit of essentially the same quality as the restricted case,
showing that current observations are consistent with both the single-sector
and two-sector realizations of the holographic entropic framework.

Notably, a joint analysis combining all datasets under $\Lambda$CDM proves
problematic due to the well-known tension between the CMB shift parameters and
the Pantheon$^+$+SH0ES compilation. In contrast, the HDE model admits regions
of parameter space in which the constraints from these datasets become mutually
compatible. This highlights an interesting phenomenological feature of the
extended holographic framework: its enlarged parameter space provides
additional flexibility in accommodating different cosmological probes while
remaining consistent with the full observational dataset.

Furthermore, we have verified that the early-time evolution of the HDE model
remains very close to that of $\Lambda$CDM, leading to negligible differences
in key quantities such as the sound horizon $r_s$ and the comoving distance
$D_M(z^*)$. This supports the use of compressed CMB observables derived within
the standard cosmological framework.

Taken together, our results show that the HDE scenario is observationally
viable and provides a consistent description of the cosmological background
evolution across a broad redshift range. While the current data do not
compellingly favor departures from $\Lambda$CDM, the model offers a
theoretically motivated and statistically acceptable alternative rooted in
generalized statistical mechanics and holographic entropy. Moreover, it
naturally encompasses the standard cosmological constant scenario as a limiting
case, while retaining the flexibility to describe a wider class of dark-energy
dynamics. In this respect, the Luciano-Saridakis framework provides a
well-motivated extension of conventional HDE models, establishing a direct link
between microscopic entropic considerations and late-time cosmological
observables.

Looking ahead, several extensions of this work deserve further investigation.
A full Boltzmann analysis incorporating the CMB temperature and polarization
power spectra, together with large-scale structure and matter power spectrum
measurements, could provide significantly tighter constraints on the entropic
parameters and reveal signatures beyond the background level. Furthermore, it 
would also be interesting to explore whether the additional flexibility of the 
generalized
entropic framework can contribute to alleviating some of the persistent
tensions in modern cosmology, such as the discrepancy in the determination of
the Hubble constant and other emerging inconsistencies between early- and
late-time probes.  These investigations are left for future projects.

\section*{Acknowledgments}
The research of GGL is supported by the postdoctoral funding program of the University of Lleida. The authors acknowledge the contribution of the LISA CosWG, and of COST Actions CA21136 “Addressing observational tensions in cosmology with systematics and fundamental physics (CosmoVerse)”, CA21106 “COSMIC WISPersin the Dark Universe: Theory, astrophysics and experiments”, and CA23130 “Bridging high and low energies in search of quantum gravity (BridgeQG)”. S.L. and M.L. are supported by grant PIP 11220200100729CO CONICET and grant 20020170100129BA UBACYT. The authors acknowledge the use of the supercluster- MIZTLI of UNAM through project LANCAD-UNAM-DGTIC-449 and thank the people of DGTIC-UNAM for technical and computational support

\printcredits

\bibliographystyle{cas-model2-names}

\bibliography{references}

\begin{thebibliography}{96}
\expandafter\ifx\csname natexlab\endcsname\relax\def\natexlab#1{#1}\fi
\providecommand{\url}[1]{\texttt{#1}}
\providecommand{\href}[2]{#2}
\providecommand{\path}[1]{#1}
\providecommand{\DOIprefix}{doi:}
\providecommand{\ArXivprefix}{arXiv:}
\providecommand{\URLprefix}{URL: }
\providecommand{\Pubmedprefix}{pmid:}
\providecommand{\doi}[1]{\href{http://dx.doi.org/#1}{\path{#1}}}
\providecommand{\Pubmed}[1]{\href{pmid:#1}{\path{#1}}}
\providecommand{\bibinfo}[2]{#2}
\ifx\xfnm\relax \def\xfnm[#1]{\unskip,\space#1}\fi
\bibitem[{Abdul~Karim et~al.(2025)}]{DESI:2025zgx}
\bibinfo{author}{Abdul~Karim, M.}, et~al. (\bibinfo{collaboration}{DESI}), \bibinfo{year}{2025}.
\newblock \bibinfo{title}{{DESI DR2 results. II. Measurements of baryon acoustic oscillations and cosmological constraints}}.
\newblock \bibinfo{journal}{Phys. Rev. D} \bibinfo{volume}{112}, \bibinfo{pages}{083515}.
\newblock \DOIprefix\doi{10.1103/tr6y-kpc6}, \href{http://arxiv.org/abs/2503.14738}{\tt arXiv:2503.14738}.
\bibitem[{Addazi et~al.(2022)}]{Addazi:2021xuf}
\bibinfo{author}{Addazi, A.}, et~al., \bibinfo{year}{2022}.
\newblock \bibinfo{title}{{Quantum gravity phenomenology at the dawn of the multi-messenger era{\textemdash}A review}}.
\newblock \bibinfo{journal}{Prog. Part. Nucl. Phys.} \bibinfo{volume}{125}, \bibinfo{pages}{103948}.
\newblock \DOIprefix\doi{10.1016/j.ppnp.2022.103948}, \href{http://arxiv.org/abs/2111.05659}{\tt arXiv:2111.05659}.
\bibitem[{Aghanim et~al.(2020)}]{Planck:2019nip}
\bibinfo{author}{Aghanim, N.}, et~al. (\bibinfo{collaboration}{Planck}), \bibinfo{year}{2020}.
\newblock \bibinfo{title}{{Planck 2018 results. V. CMB power spectra and likelihoods}}.
\newblock \bibinfo{journal}{Astron. Astrophys.} \bibinfo{volume}{641}, \bibinfo{pages}{A5}.
\newblock \DOIprefix\doi{10.1051/0004-6361/201936386}, \href{http://arxiv.org/abs/1907.12875}{\tt arXiv:1907.12875}.
\bibitem[{Akaike(1974)}]{Akaike:1974vps}
\bibinfo{author}{Akaike, H.}, \bibinfo{year}{1974}.
\newblock \bibinfo{title}{{A new look at the statistical model identification}}.
\newblock \bibinfo{journal}{IEEE Trans. Automatic Control} \bibinfo{volume}{19}, \bibinfo{pages}{716--723}.
\newblock \DOIprefix\doi{10.1109/TAC.1974.1100705}.
\bibitem[{Anagnostopoulos et~al.(2020)Anagnostopoulos, Basilakos and Saridakis}]{Anagnostopoulos:2020ctz}
\bibinfo{author}{Anagnostopoulos, F.K.}, \bibinfo{author}{Basilakos, S.}, \bibinfo{author}{Saridakis, E.N.}, \bibinfo{year}{2020}.
\newblock \bibinfo{title}{{Observational constraints on Barrow holographic dark energy}}.
\newblock \bibinfo{journal}{Eur. Phys. J. C} \bibinfo{volume}{80}, \bibinfo{pages}{826}.
\newblock \DOIprefix\doi{10.1140/epjc/s10052-020-8360-5}, \href{http://arxiv.org/abs/2005.10302}{\tt arXiv:2005.10302}.
\bibitem[{Aviles et~al.(2011)Aviles, Bonanno, Luongo and Quevedo}]{Aviles:2011sfa}
\bibinfo{author}{Aviles, A.}, \bibinfo{author}{Bonanno, L.}, \bibinfo{author}{Luongo, O.}, \bibinfo{author}{Quevedo, H.}, \bibinfo{year}{2011}.
\newblock \bibinfo{title}{{Holographic dark matter and dark energy with second order invariants}}.
\newblock \bibinfo{journal}{Phys. Rev. D} \bibinfo{volume}{84}, \bibinfo{pages}{103520}.
\newblock \DOIprefix\doi{10.1103/PhysRevD.84.103520}, \href{http://arxiv.org/abs/1109.3177}{\tt arXiv:1109.3177}.
\bibitem[{Bak and Rey(2000)}]{Bak:1999hd}
\bibinfo{author}{Bak, D.}, \bibinfo{author}{Rey, S.J.}, \bibinfo{year}{2000}.
\newblock \bibinfo{title}{{Cosmic holography}}.
\newblock \bibinfo{journal}{Class. Quant. Grav.} \bibinfo{volume}{17}, \bibinfo{pages}{L83}.
\newblock \DOIprefix\doi{10.1088/0264-9381/17/15/101}, \href{http://arxiv.org/abs/hep-th/9902173}{\tt arXiv:hep-th/9902173}.
\bibitem[{Bamba et~al.(2012)Bamba, Capozziello, Nojiri and Odintsov}]{Bamba:2012cp}
\bibinfo{author}{Bamba, K.}, \bibinfo{author}{Capozziello, S.}, \bibinfo{author}{Nojiri, S.}, \bibinfo{author}{Odintsov, S.D.}, \bibinfo{year}{2012}.
\newblock \bibinfo{title}{{Dark energy cosmology: the equivalent description via different theoretical models and cosmography tests}}.
\newblock \bibinfo{journal}{Astrophys. Space Sci.} \bibinfo{volume}{342}, \bibinfo{pages}{155--228}.
\newblock \DOIprefix\doi{10.1007/s10509-012-1181-8}, \href{http://arxiv.org/abs/1205.3421}{\tt arXiv:1205.3421}.
\bibitem[{Barrow(2020)}]{Barrow:2020tzx}
\bibinfo{author}{Barrow, J.D.}, \bibinfo{year}{2020}.
\newblock \bibinfo{title}{{The Area of a Rough Black Hole}}.
\newblock \bibinfo{journal}{Phys. Lett. B} \bibinfo{volume}{808}, \bibinfo{pages}{135643}.
\newblock \DOIprefix\doi{10.1016/j.physletb.2020.135643}, \href{http://arxiv.org/abs/2004.09444}{\tt arXiv:2004.09444}.
\bibitem[{Bhattacharjee(2021)}]{Bhattacharjee:2020ixg}
\bibinfo{author}{Bhattacharjee, S.}, \bibinfo{year}{2021}.
\newblock \bibinfo{title}{{Growth Rate and Configurational Entropy in Tsallis Holographic Dark Energy}}.
\newblock \bibinfo{journal}{Eur. Phys. J. C} \bibinfo{volume}{81}, \bibinfo{pages}{217}.
\newblock \DOIprefix\doi{10.1140/epjc/s10052-021-09003-0}, \href{http://arxiv.org/abs/2011.13135}{\tt arXiv:2011.13135}.
\bibitem[{{Blas} et~al.(2011){Blas}, {Lesgourgues} and {Tram}}]{2011JCAP...07..034B}
\bibinfo{author}{{Blas}, D.}, \bibinfo{author}{{Lesgourgues}, J.}, \bibinfo{author}{{Tram}, T.}, \bibinfo{year}{2011}.
\newblock \bibinfo{title}{{The Cosmic Linear Anisotropy Solving System (CLASS). Part II: Approximation schemes}}.
\newblock \bibinfo{journal}{\jcap} \bibinfo{volume}{2011}, \bibinfo{pages}{034}.
\newblock \DOIprefix\doi{10.1088/1475-7516/2011/07/034}, \href{http://arxiv.org/abs/1104.2933}{\tt arXiv:1104.2933}.
\bibitem[{Bouhmadi-Lopez et~al.(2011)Bouhmadi-Lopez, Errahmani and Ouali}]{Bouhmadi-Lopez:2011qvd}
\bibinfo{author}{Bouhmadi-Lopez, M.}, \bibinfo{author}{Errahmani, A.}, \bibinfo{author}{Ouali, T.}, \bibinfo{year}{2011}.
\newblock \bibinfo{title}{{The cosmology of an holographic induced gravity model with curvature effects}}.
\newblock \bibinfo{journal}{Phys. Rev. D} \bibinfo{volume}{84}, \bibinfo{pages}{083508}.
\newblock \DOIprefix\doi{10.1103/PhysRevD.84.083508}, \href{http://arxiv.org/abs/1104.1181}{\tt arXiv:1104.1181}.
\bibitem[{Bousso(2002)}]{Bousso:2002ju}
\bibinfo{author}{Bousso, R.}, \bibinfo{year}{2002}.
\newblock \bibinfo{title}{{The Holographic principle}}.
\newblock \bibinfo{journal}{Rev. Mod. Phys.} \bibinfo{volume}{74}, \bibinfo{pages}{825--874}.
\newblock \DOIprefix\doi{10.1103/RevModPhys.74.825}, \href{http://arxiv.org/abs/hep-th/0203101}{\tt arXiv:hep-th/0203101}.
\bibitem[{Brout et~al.(2022)}]{Brout:2022vxf}
\bibinfo{author}{Brout, D.}, et~al., \bibinfo{year}{2022}.
\newblock \bibinfo{title}{{The Pantheon+ Analysis: Cosmological Constraints}}.
\newblock \bibinfo{journal}{Astrophys. J.} \bibinfo{volume}{938}, \bibinfo{pages}{110}.
\newblock \DOIprefix\doi{10.3847/1538-4357/ac8e04}, \href{http://arxiv.org/abs/2202.04077}{\tt arXiv:2202.04077}.
\bibitem[{Cai(2007)}]{Cai:2007us}
\bibinfo{author}{Cai, R.G.}, \bibinfo{year}{2007}.
\newblock \bibinfo{title}{{A Dark Energy Model Characterized by the Age of the Universe}}.
\newblock \bibinfo{journal}{Phys. Lett. B} \bibinfo{volume}{657}, \bibinfo{pages}{228--231}.
\newblock \DOIprefix\doi{10.1016/j.physletb.2007.09.061}, \href{http://arxiv.org/abs/0707.4049}{\tt arXiv:0707.4049}.
\bibitem[{Cai et~al.(2016)Cai, Capozziello, De~Laurentis and Saridakis}]{Cai:2015emx}
\bibinfo{author}{Cai, Y.F.}, \bibinfo{author}{Capozziello, S.}, \bibinfo{author}{De~Laurentis, M.}, \bibinfo{author}{Saridakis, E.N.}, \bibinfo{year}{2016}.
\newblock \bibinfo{title}{{f(T) teleparallel gravity and cosmology}}.
\newblock \bibinfo{journal}{Rept. Prog. Phys.} \bibinfo{volume}{79}, \bibinfo{pages}{106901}.
\newblock \DOIprefix\doi{10.1088/0034-4885/79/10/106901}, \href{http://arxiv.org/abs/1511.07586}{\tt arXiv:1511.07586}.
\bibitem[{Cai et~al.(2010)Cai, Saridakis, Setare and Xia}]{Cai:2009zp}
\bibinfo{author}{Cai, Y.F.}, \bibinfo{author}{Saridakis, E.N.}, \bibinfo{author}{Setare, M.R.}, \bibinfo{author}{Xia, J.Q.}, \bibinfo{year}{2010}.
\newblock \bibinfo{title}{{Quintom Cosmology: Theoretical implications and observations}}.
\newblock \bibinfo{journal}{Phys. Rept.} \bibinfo{volume}{493}, \bibinfo{pages}{1}.
\newblock \DOIprefix\doi{10.1016/j.physrep.2010.04.001}, \href{http://arxiv.org/abs/0909.2776}{\tt arXiv:0909.2776}.
\bibitem[{Capozziello and De~Laurentis(2011)}]{Capozziello:2011et}
\bibinfo{author}{Capozziello, S.}, \bibinfo{author}{De~Laurentis, M.}, \bibinfo{year}{2011}.
\newblock \bibinfo{title}{{Extended Theories of Gravity}}.
\newblock \bibinfo{journal}{Phys. Rept.} \bibinfo{volume}{509}, \bibinfo{pages}{167--321}.
\newblock \DOIprefix\doi{10.1016/j.physrep.2011.09.003}, \href{http://arxiv.org/abs/1108.6266}{\tt arXiv:1108.6266}.
\bibitem[{Chantada et~al.(2023)Chantada, Landau, Protopapas, Sc\'occola and Garraffo}]{Chantada2023}
\bibinfo{author}{Chantada, A.T.}, \bibinfo{author}{Landau, S.J.}, \bibinfo{author}{Protopapas, P.}, \bibinfo{author}{Sc\'occola, C.G.}, \bibinfo{author}{Garraffo, C.}, \bibinfo{year}{2023}.
\newblock \bibinfo{title}{Cosmology-informed neural networks to solve the background dynamics of the {Universe}}.
\newblock \bibinfo{journal}{Phys. Rev. D} \bibinfo{volume}{107}, \bibinfo{pages}{063523}.
\newblock \DOIprefix\doi{10.1103/PhysRevD.107.063523}.
\bibitem[{Cimdiker et~al.(2025)Cimdiker, Dabrowski and Salzano}]{Cimdiker:2025vfn}
\bibinfo{author}{Cimdiker, I.}, \bibinfo{author}{Dabrowski, M.P.}, \bibinfo{author}{Salzano, V.}, \bibinfo{year}{2025}.
\newblock \bibinfo{title}{{Generalized nonextensive entropy holographic dark energy models verified by cosmological data}}.
\newblock \bibinfo{journal}{Eur. Phys. J. C} \bibinfo{volume}{85}, \bibinfo{pages}{775}.
\newblock \DOIprefix\doi{10.1140/epjc/s10052-025-14498-y}, \href{http://arxiv.org/abs/2503.18230}{\tt arXiv:2503.18230}.
\bibitem[{Cohen et~al.(1999)Cohen, Kaplan and Nelson}]{Cohen:1998zx}
\bibinfo{author}{Cohen, A.G.}, \bibinfo{author}{Kaplan, D.B.}, \bibinfo{author}{Nelson, A.E.}, \bibinfo{year}{1999}.
\newblock \bibinfo{title}{{Effective field theory, black holes, and the cosmological constant}}.
\newblock \bibinfo{journal}{Phys. Rev. Lett.} \bibinfo{volume}{82}, \bibinfo{pages}{4971--4974}.
\newblock \DOIprefix\doi{10.1103/PhysRevLett.82.4971}, \href{http://arxiv.org/abs/hep-th/9803132}{\tt arXiv:hep-th/9803132}.
\bibitem[{Copeland et~al.(2006)Copeland, Sami and Tsujikawa}]{Copeland:2006wr}
\bibinfo{author}{Copeland, E.J.}, \bibinfo{author}{Sami, M.}, \bibinfo{author}{Tsujikawa, S.}, \bibinfo{year}{2006}.
\newblock \bibinfo{title}{{Dynamics of dark energy}}.
\newblock \bibinfo{journal}{Int. J. Mod. Phys. D} \bibinfo{volume}{15}, \bibinfo{pages}{1753}.
\newblock \DOIprefix\doi{10.1142/S021827180600942X}, \href{http://arxiv.org/abs/hep-th/0603057}{\tt arXiv:hep-th/0603057}.
\bibitem[{Dabrowski and Salzano(2020)}]{Dabrowski:2020atl}
\bibinfo{author}{Dabrowski, M.P.}, \bibinfo{author}{Salzano, V.}, \bibinfo{year}{2020}.
\newblock \bibinfo{title}{{Geometrical observational bounds on a fractal horizon holographic dark energy}}.
\newblock \bibinfo{journal}{Phys. Rev. D} \bibinfo{volume}{102}, \bibinfo{pages}{064047}.
\newblock \DOIprefix\doi{10.1103/PhysRevD.102.064047}, \href{http://arxiv.org/abs/2009.08306}{\tt arXiv:2009.08306}.
\bibitem[{D'Agostino(2019)}]{DAgostino:2019wko}
\bibinfo{author}{D'Agostino, R.}, \bibinfo{year}{2019}.
\newblock \bibinfo{title}{{Holographic dark energy from nonadditive entropy: cosmological perturbations and observational constraints}}.
\newblock \bibinfo{journal}{Phys. Rev. D} \bibinfo{volume}{99}, \bibinfo{pages}{103524}.
\newblock \DOIprefix\doi{10.1103/PhysRevD.99.103524}, \href{http://arxiv.org/abs/1903.03836}{\tt arXiv:1903.03836}.
\bibitem[{Di~Valentino et~al.(2025)}]{CosmoVerseNetwork:2025alb}
\bibinfo{author}{Di~Valentino, E.}, et~al. (\bibinfo{collaboration}{CosmoVerse Network}), \bibinfo{year}{2025}.
\newblock \bibinfo{title}{{The CosmoVerse White Paper: Addressing observational tensions in cosmology with systematics and fundamental physics}}.
\newblock \bibinfo{journal}{Phys. Dark Univ.} \bibinfo{volume}{49}, \bibinfo{pages}{101965}.
\newblock \DOIprefix\doi{10.1016/j.dark.2025.101965}, \href{http://arxiv.org/abs/2504.01669}{\tt arXiv:2504.01669}.
\bibitem[{Drepanou et~al.(2022)Drepanou, Lymperis, Saridakis and Yesmakhanova}]{Drepanou:2021jiv}
\bibinfo{author}{Drepanou, N.}, \bibinfo{author}{Lymperis, A.}, \bibinfo{author}{Saridakis, E.N.}, \bibinfo{author}{Yesmakhanova, K.}, \bibinfo{year}{2022}.
\newblock \bibinfo{title}{{Kaniadakis holographic dark energy and cosmology}}.
\newblock \bibinfo{journal}{Eur. Phys. J. C} \bibinfo{volume}{82}, \bibinfo{pages}{449}.
\newblock \DOIprefix\doi{10.1140/epjc/s10052-022-10415-9}, \href{http://arxiv.org/abs/2109.09181}{\tt arXiv:2109.09181}.
\bibitem[{Duran et~al.(2010)Duran, Pavon and Zimdahl}]{Duran:2010hi}
\bibinfo{author}{Duran, I.}, \bibinfo{author}{Pavon, D.}, \bibinfo{author}{Zimdahl, W.}, \bibinfo{year}{2010}.
\newblock \bibinfo{title}{{Observational constraints on a holographic, interacting dark energy model}}.
\newblock \bibinfo{journal}{JCAP} \bibinfo{volume}{07}, \bibinfo{pages}{018}.
\newblock \DOIprefix\doi{10.1088/1475-7516/2010/07/018}, \href{http://arxiv.org/abs/1007.0390}{\tt arXiv:1007.0390}.
\bibitem[{Fischler and Susskind(1998)}]{Fischler:1998st}
\bibinfo{author}{Fischler, W.}, \bibinfo{author}{Susskind, L.}, \bibinfo{year}{1998}.
\newblock \bibinfo{title}{Holography and cosmology} \href{http://arxiv.org/abs/hep-th/9806039}{\tt arXiv:hep-th/9806039}.
\bibitem[{Ghaffari et~al.(2023)Ghaffari, Luciano and Capozziello}]{Ghaffari:2022skp}
\bibinfo{author}{Ghaffari, S.}, \bibinfo{author}{Luciano, G.G.}, \bibinfo{author}{Capozziello, S.}, \bibinfo{year}{2023}.
\newblock \bibinfo{title}{{Barrow holographic dark energy in the Brans{\textendash}Dicke cosmology}}.
\newblock \bibinfo{journal}{Eur. Phys. J. Plus} \bibinfo{volume}{138}, \bibinfo{pages}{82}.
\newblock \DOIprefix\doi{10.1140/epjp/s13360-022-03481-1}, \href{http://arxiv.org/abs/2209.00903}{\tt arXiv:2209.00903}.
\bibitem[{Goldstein et~al.(2020)Goldstein, Lebowitz, Tumulka and Zangh{\`\i}}]{goldstein2020gibbs}
\bibinfo{author}{Goldstein, S.}, \bibinfo{author}{Lebowitz, J.L.}, \bibinfo{author}{Tumulka, R.}, \bibinfo{author}{Zangh{\`\i}, N.}, \bibinfo{year}{2020}.
\newblock \bibinfo{title}{Gibbs and boltzmann entropy in classical and quantum mechanics}, in: \bibinfo{booktitle}{Statistical mechanics and scientific explanation: Determinism, indeterminism and laws of nature}. \bibinfo{publisher}{World Scientific}, pp. \bibinfo{pages}{519--581}.
\bibitem[{Gong(2004)}]{Gong:2004fq}
\bibinfo{author}{Gong, Y.}, \bibinfo{year}{2004}.
\newblock \bibinfo{title}{{Extended holographic dark energy}}.
\newblock \bibinfo{journal}{Phys. Rev. D} \bibinfo{volume}{70}, \bibinfo{pages}{064029}.
\newblock \DOIprefix\doi{10.1103/PhysRevD.70.064029}, \href{http://arxiv.org/abs/hep-th/0404030}{\tt arXiv:hep-th/0404030}.
\bibitem[{Gong and Li(2010)}]{Gong:2009dc}
\bibinfo{author}{Gong, Y.}, \bibinfo{author}{Li, T.}, \bibinfo{year}{2010}.
\newblock \bibinfo{title}{{A Modified Holographic Dark Energy Model with Infrared Infinite Extra Dimension(s)}}.
\newblock \bibinfo{journal}{Phys. Lett. B} \bibinfo{volume}{683}, \bibinfo{pages}{241--247}.
\newblock \DOIprefix\doi{10.1016/j.physletb.2009.12.040}, \href{http://arxiv.org/abs/0907.0860}{\tt arXiv:0907.0860}.
\bibitem[{Hanel and Thurner(2011)}]{hanel2011comprehensive}
\bibinfo{author}{Hanel, R.}, \bibinfo{author}{Thurner, S.}, \bibinfo{year}{2011}.
\newblock \bibinfo{title}{A comprehensive classification of complex statistical systems and an axiomatic derivation of their entropy and distribution functions}.
\newblock \bibinfo{journal}{Europhysics Letters} \bibinfo{volume}{93}, \bibinfo{pages}{20006}.
\bibitem[{Hern\'andez-Almada et~al.(2022)Hern\'andez-Almada, Leon, Maga\~na, Garc\'\i{}a-Aspeitia, Motta, Saridakis, Yesmakhanova and Millano}]{Hernandez-Almada:2021rjs}
\bibinfo{author}{Hern\'andez-Almada, A.}, \bibinfo{author}{Leon, G.}, \bibinfo{author}{Maga\~na, J.}, \bibinfo{author}{Garc\'\i{}a-Aspeitia, M.A.}, \bibinfo{author}{Motta, V.}, \bibinfo{author}{Saridakis, E.N.}, \bibinfo{author}{Yesmakhanova, K.}, \bibinfo{author}{Millano, A.D.}, \bibinfo{year}{2022}.
\newblock \bibinfo{title}{{Observational constraints and dynamical analysis of Kaniadakis horizon-entropy cosmology}}.
\newblock \bibinfo{journal}{Mon. Not. Roy. Astron. Soc.} \bibinfo{volume}{512}, \bibinfo{pages}{5122--5134}.
\newblock \DOIprefix\doi{10.1093/mnras/stac795}, \href{http://arxiv.org/abs/2112.04615}{\tt arXiv:2112.04615}.
\bibitem[{Hern{\'a}ndez-Almada et~al.(2022)Hern{\'a}ndez-Almada, Leon, Maga{\~n}a, Garc{\'\i}a-Aspeitia, Motta, Saridakis and Yesmakhanova}]{Hernandez-Almada:2021aiw}
\bibinfo{author}{Hern{\'a}ndez-Almada, A.}, \bibinfo{author}{Leon, G.}, \bibinfo{author}{Maga{\~n}a, J.}, \bibinfo{author}{Garc{\'\i}a-Aspeitia, M.A.}, \bibinfo{author}{Motta, V.}, \bibinfo{author}{Saridakis, E.N.}, \bibinfo{author}{Yesmakhanova, K.}, \bibinfo{year}{2022}.
\newblock \bibinfo{title}{{Kaniadakis-holographic dark energy: observational constraints and global dynamics}}.
\newblock \bibinfo{journal}{Mon. Not. Roy. Astron. Soc.} \bibinfo{volume}{511}, \bibinfo{pages}{4147--4158}.
\newblock \DOIprefix\doi{10.1093/mnras/stac255}, \href{http://arxiv.org/abs/2111.00558}{\tt arXiv:2111.00558}.
\bibitem[{'t~Hooft(1993)}]{tHooft:1993dmi}
\bibinfo{author}{'t~Hooft, G.}, \bibinfo{year}{1993}.
\newblock \bibinfo{title}{{Dimensional reduction in quantum gravity}}.
\newblock \bibinfo{journal}{Conf. Proc. C} \bibinfo{volume}{930308}, \bibinfo{pages}{284--296}.
\newblock \href{http://arxiv.org/abs/gr-qc/9310026}{\tt arXiv:gr-qc/9310026}.
\bibitem[{Horava and Minic(2000)}]{Horava:2000tb}
\bibinfo{author}{Horava, P.}, \bibinfo{author}{Minic, D.}, \bibinfo{year}{2000}.
\newblock \bibinfo{title}{{Probable values of the cosmological constant in a holographic theory}}.
\newblock \bibinfo{journal}{Phys. Rev. Lett.} \bibinfo{volume}{85}, \bibinfo{pages}{1610--1613}.
\newblock \DOIprefix\doi{10.1103/PhysRevLett.85.1610}, \href{http://arxiv.org/abs/hep-th/0001145}{\tt arXiv:hep-th/0001145}.
\bibitem[{Huang et~al.(2021)Huang, Huang, Xu, Tu and Chen}]{Huang:2021zgj}
\bibinfo{author}{Huang, Q.}, \bibinfo{author}{Huang, H.}, \bibinfo{author}{Xu, B.}, \bibinfo{author}{Tu, F.}, \bibinfo{author}{Chen, J.}, \bibinfo{year}{2021}.
\newblock \bibinfo{title}{{Dynamical analysis and statefinder of Barrow holographic dark energy}}.
\newblock \bibinfo{journal}{Eur. Phys. J. C} \bibinfo{volume}{81}, \bibinfo{pages}{686}.
\newblock \DOIprefix\doi{10.1140/epjc/s10052-021-09480-3}, \href{http://arxiv.org/abs/2201.11414}{\tt arXiv:2201.11414}.
\bibitem[{Huang and Li(2004)}]{Huang:2004ai}
\bibinfo{author}{Huang, Q.G.}, \bibinfo{author}{Li, M.}, \bibinfo{year}{2004}.
\newblock \bibinfo{title}{{The Holographic dark energy in a non-flat universe}}.
\newblock \bibinfo{journal}{JCAP} \bibinfo{volume}{08}, \bibinfo{pages}{013}.
\newblock \DOIprefix\doi{10.1088/1475-7516/2004/08/013}, \href{http://arxiv.org/abs/astro-ph/0404229}{\tt arXiv:astro-ph/0404229}.
\bibitem[{Jamil and Saridakis(2010)}]{Jamil:2010vr}
\bibinfo{author}{Jamil, M.}, \bibinfo{author}{Saridakis, E.N.}, \bibinfo{year}{2010}.
\newblock \bibinfo{title}{{New agegraphic dark energy in Horava-Lifshitz cosmology}}.
\newblock \bibinfo{journal}{JCAP} \bibinfo{volume}{07}, \bibinfo{pages}{028}.
\newblock \DOIprefix\doi{10.1088/1475-7516/2010/07/028}, \href{http://arxiv.org/abs/1003.5637}{\tt arXiv:1003.5637}.
\bibitem[{Jusufi et~al.(2022)Jusufi, Azreg-A\"\i{}nou, Jamil and Saridakis}]{Jusufi:2021fek}
\bibinfo{author}{Jusufi, K.}, \bibinfo{author}{Azreg-A\"\i{}nou, M.}, \bibinfo{author}{Jamil, M.}, \bibinfo{author}{Saridakis, E.N.}, \bibinfo{year}{2022}.
\newblock \bibinfo{title}{{Constraints on Barrow Entropy from M87* and S2 Star Observations}}.
\newblock \bibinfo{journal}{Universe} \bibinfo{volume}{8}, \bibinfo{pages}{102}.
\newblock \DOIprefix\doi{10.3390/universe8020102}, \href{http://arxiv.org/abs/2110.07258}{\tt arXiv:2110.07258}.
\bibitem[{Khinchin(1957)}]{Khinchin1957}
\bibinfo{author}{Khinchin, A.}, \bibinfo{year}{1957}.
\newblock \bibinfo{title}{Mathematical Foundations of Information Theory}.
\newblock \bibinfo{publisher}{Dover Publications}, \bibinfo{address}{New York, NY, USA}.
\bibitem[{Kim et~al.(2006)Kim, Lee and Myung}]{Kim:2005at}
\bibinfo{author}{Kim, H.}, \bibinfo{author}{Lee, H.W.}, \bibinfo{author}{Myung, Y.S.}, \bibinfo{year}{2006}.
\newblock \bibinfo{title}{{Equation of state for an interacting holographic dark energy model}}.
\newblock \bibinfo{journal}{Phys. Lett. B} \bibinfo{volume}{632}, \bibinfo{pages}{605--609}.
\newblock \DOIprefix\doi{10.1016/j.physletb.2005.11.043}, \href{http://arxiv.org/abs/gr-qc/0509040}{\tt arXiv:gr-qc/0509040}.
\bibitem[{Landim(2016)}]{Landim:2015hqa}
\bibinfo{author}{Landim, R.C.G.}, \bibinfo{year}{2016}.
\newblock \bibinfo{title}{{Holographic dark energy from minimal supergravity}}.
\newblock \bibinfo{journal}{Int. J. Mod. Phys. D} \bibinfo{volume}{25}, \bibinfo{pages}{1650050}.
\newblock \DOIprefix\doi{10.1142/S0218271816500504}, \href{http://arxiv.org/abs/1508.07248}{\tt arXiv:1508.07248}.
\bibitem[{Landim(2022)}]{Landim:2022jgr}
\bibinfo{author}{Landim, R.G.}, \bibinfo{year}{2022}.
\newblock \bibinfo{title}{{Note on interacting holographic dark energy with a Hubble-scale cutoff}}.
\newblock \bibinfo{journal}{Phys. Rev. D} \bibinfo{volume}{106}, \bibinfo{pages}{043527}.
\newblock \DOIprefix\doi{10.1103/PhysRevD.106.043527}, \href{http://arxiv.org/abs/2206.10205}{\tt arXiv:2206.10205}.
\bibitem[{Li and Wang(2025)}]{Li:2024bwr}
\bibinfo{author}{Li, J.X.}, \bibinfo{author}{Wang, S.}, \bibinfo{year}{2025}.
\newblock \bibinfo{title}{{A comprehensive numerical study on four categories of holographic dark energy models}}.
\newblock \bibinfo{journal}{JCAP} \bibinfo{volume}{07}, \bibinfo{pages}{047}.
\newblock \DOIprefix\doi{10.1088/1475-7516/2025/07/047}, \href{http://arxiv.org/abs/2412.09064}{\tt arXiv:2412.09064}.
\bibitem[{Li(2004)}]{Li:2004rb}
\bibinfo{author}{Li, M.}, \bibinfo{year}{2004}.
\newblock \bibinfo{title}{{A Model of holographic dark energy}}.
\newblock \bibinfo{journal}{Phys. Lett. B} \bibinfo{volume}{603}, \bibinfo{pages}{1}.
\newblock \DOIprefix\doi{10.1016/j.physletb.2004.10.014}, \href{http://arxiv.org/abs/hep-th/0403127}{\tt arXiv:hep-th/0403127}.
\bibitem[{Li et~al.(2009)Li, Li, Wang and Zhang}]{Li:2009bn}
\bibinfo{author}{Li, M.}, \bibinfo{author}{Li, X.D.}, \bibinfo{author}{Wang, S.}, \bibinfo{author}{Zhang, X.}, \bibinfo{year}{2009}.
\newblock \bibinfo{title}{{Holographic dark energy models: A comparison from the latest observational data}}.
\newblock \bibinfo{journal}{JCAP} \bibinfo{volume}{06}, \bibinfo{pages}{036}.
\newblock \DOIprefix\doi{10.1088/1475-7516/2009/06/036}, \href{http://arxiv.org/abs/0904.0928}{\tt arXiv:0904.0928}.
\bibitem[{{Liu} et~al.(2024){Liu}, {Wang} and {Zhao}}]{2024PhLB..85438717L}
\bibinfo{author}{{Liu}, G.}, \bibinfo{author}{{Wang}, Y.}, \bibinfo{author}{{Zhao}, W.}, \bibinfo{year}{2024}.
\newblock \bibinfo{title}{{Testing the consistency of early and late cosmological parameters with BAO and CMB data}}.
\newblock \bibinfo{journal}{Physics Letters B} \bibinfo{volume}{854}, \bibinfo{pages}{138717}.
\newblock \DOIprefix\doi{10.1016/j.physletb.2024.138717}, \href{http://arxiv.org/abs/2401.10571}{\tt arXiv:2401.10571}.
\bibitem[{Lu et~al.(2010)Lu, Saridakis, Setare and Xu}]{Lu:2009iv}
\bibinfo{author}{Lu, J.}, \bibinfo{author}{Saridakis, E.N.}, \bibinfo{author}{Setare, M.R.}, \bibinfo{author}{Xu, L.}, \bibinfo{year}{2010}.
\newblock \bibinfo{title}{{Observational constraints on holographic dark energy with varying gravitational constant}}.
\newblock \bibinfo{journal}{JCAP} \bibinfo{volume}{03}, \bibinfo{pages}{031}.
\newblock \DOIprefix\doi{10.1088/1475-7516/2010/03/031}, \href{http://arxiv.org/abs/0912.0923}{\tt arXiv:0912.0923}.
\bibitem[{Luciano(2022)}]{Luciano:2022ffn}
\bibinfo{author}{Luciano, G.G.}, \bibinfo{year}{2022}.
\newblock \bibinfo{title}{{Cosmic evolution and thermal stability of Barrow holographic dark energy in a nonflat Friedmann-Robertson-Walker Universe}}.
\newblock \bibinfo{journal}{Phys. Rev. D} \bibinfo{volume}{106}, \bibinfo{pages}{083530}.
\newblock \DOIprefix\doi{10.1103/PhysRevD.106.083530}, \href{http://arxiv.org/abs/2210.06320}{\tt arXiv:2210.06320}.
\bibitem[{Luciano(2025)}]{Luciano:2025fox}
\bibinfo{author}{Luciano, G.G.}, \bibinfo{year}{2025}.
\newblock \bibinfo{title}{{Quantum Entropy-Driven Modifications to Holographic Dark Energy in $f(G,T)$ Gravity}}.
\newblock \bibinfo{journal}{Eur. Phys. J. C} \bibinfo{volume}{85}, \bibinfo{pages}{572}.
\newblock \DOIprefix\doi{10.1140/epjc/s10052-025-14272-0}, \href{http://arxiv.org/abs/2503.22609}{\tt arXiv:2503.22609}.
\bibitem[{Luciano and Gin{\'e}(2023)}]{Luciano:2022hhy}
\bibinfo{author}{Luciano, G.G.}, \bibinfo{author}{Gin{\'e}, J.}, \bibinfo{year}{2023}.
\newblock \bibinfo{title}{{Generalized interacting Barrow Holographic Dark Energy: Cosmological predictions and thermodynamic considerations}}.
\newblock \bibinfo{journal}{Phys. Dark Univ.} \bibinfo{volume}{41}, \bibinfo{pages}{101256}.
\newblock \DOIprefix\doi{10.1016/j.dark.2023.101256}, \href{http://arxiv.org/abs/2210.09755}{\tt arXiv:2210.09755}.
\bibitem[{Luciano and Saridakis(2022)}]{Luciano:2022pzg}
\bibinfo{author}{Luciano, G.G.}, \bibinfo{author}{Saridakis, E.N.}, \bibinfo{year}{2022}.
\newblock \bibinfo{title}{{Baryon asymmetry from Barrow entropy: theoretical predictions and observational constraints}}.
\newblock \bibinfo{journal}{Eur. Phys. J. C} \bibinfo{volume}{82}, \bibinfo{pages}{558}.
\newblock \DOIprefix\doi{10.1140/epjc/s10052-022-10530-7}, \href{http://arxiv.org/abs/2203.12010}{\tt arXiv:2203.12010}.
\bibitem[{Luciano and Saridakis(2026a)}]{Luciano:2026eiy}
\bibinfo{author}{Luciano, G.G.}, \bibinfo{author}{Saridakis, E.N.}, \bibinfo{year}{2026}a.
\newblock \bibinfo{title}{{Holographic dark energy from a new two-parameter entropic functional}}.
\newblock \bibinfo{journal}{Phys. Lett. B} \bibinfo{volume}{879}, \bibinfo{pages}{140674}.
\newblock \DOIprefix\doi{10.1016/j.physletb.2026.140674}, \href{http://arxiv.org/abs/2603.10964}{\tt arXiv:2603.10964}.
\bibitem[{Luciano and Saridakis(2026b)}]{Luciano:2026ufu}
\bibinfo{author}{Luciano, G.G.}, \bibinfo{author}{Saridakis, E.N.}, \bibinfo{year}{2026}b.
\newblock \bibinfo{title}{{New modified cosmology from a new generalized entropy}}.
\newblock \bibinfo{journal}{Phys. Lett. B} \bibinfo{volume}{879}, \bibinfo{pages}{140703}.
\newblock \DOIprefix\doi{10.1016/j.physletb.2026.140703}, \href{http://arxiv.org/abs/2602.20004}{\tt arXiv:2602.20004}.
\bibitem[{Micheletti(2010)}]{Micheletti:2009jy}
\bibinfo{author}{Micheletti, S.M.R.}, \bibinfo{year}{2010}.
\newblock \bibinfo{title}{{Observational constraints on holographic tachyonic dark energy in interaction with dark matter}}.
\newblock \bibinfo{journal}{JCAP} \bibinfo{volume}{05}, \bibinfo{pages}{009}.
\newblock \DOIprefix\doi{10.1088/1475-7516/2010/05/009}, \href{http://arxiv.org/abs/0912.3992}{\tt arXiv:0912.3992}.
\bibitem[{Moresco(2015)}]{Moresco:2015cya}
\bibinfo{author}{Moresco, M.}, \bibinfo{year}{2015}.
\newblock \bibinfo{title}{{Raising the bar: new constraints on the Hubble parameter with cosmic chronometers at z {\ensuremath{\sim}} 2}}.
\newblock \bibinfo{journal}{Mon. Not. Roy. Astron. Soc.} \bibinfo{volume}{450}, \bibinfo{pages}{L16--L20}.
\newblock \DOIprefix\doi{10.1093/mnrasl/slv037}, \href{http://arxiv.org/abs/1503.01116}{\tt arXiv:1503.01116}.
\bibitem[{{Moresco} et~al.(2020){Moresco}, {Jimenez}, {Verde}, {Cimatti} and {Pozzetti}}]{2020ApJ...898...82M}
\bibinfo{author}{{Moresco}, M.}, \bibinfo{author}{{Jimenez}, R.}, \bibinfo{author}{{Verde}, L.}, \bibinfo{author}{{Cimatti}, A.}, \bibinfo{author}{{Pozzetti}, L.}, \bibinfo{year}{2020}.
\newblock \bibinfo{title}{{Setting the Stage for Cosmic Chronometers. II. Impact of Stellar Population Synthesis Models Systematics and Full Covariance Matrix}}.
\newblock \bibinfo{journal}{\apj} \bibinfo{volume}{898}, \bibinfo{pages}{82}.
\newblock \DOIprefix\doi{10.3847/1538-4357/ab9eb0}, \href{http://arxiv.org/abs/2003.07362}{\tt arXiv:2003.07362}.
\bibitem[{Moresco et~al.(2016)Moresco, Pozzetti, Cimatti, Jimenez, Maraston, Verde, Thomas, Citro, Tojeiro and Wilkinson}]{Moresco:2016mzx}
\bibinfo{author}{Moresco, M.}, \bibinfo{author}{Pozzetti, L.}, \bibinfo{author}{Cimatti, A.}, \bibinfo{author}{Jimenez, R.}, \bibinfo{author}{Maraston, C.}, \bibinfo{author}{Verde, L.}, \bibinfo{author}{Thomas, D.}, \bibinfo{author}{Citro, A.}, \bibinfo{author}{Tojeiro, R.}, \bibinfo{author}{Wilkinson, D.}, \bibinfo{year}{2016}.
\newblock \bibinfo{title}{{A 6{\%} measurement of the Hubble parameter at $z\sim0.45$: direct evidence of the epoch of cosmic re-acceleration}}.
\newblock \bibinfo{journal}{JCAP} \bibinfo{volume}{05}, \bibinfo{pages}{014}.
\newblock \DOIprefix\doi{10.1088/1475-7516/2016/05/014}, \href{http://arxiv.org/abs/1601.01701}{\tt arXiv:1601.01701}.
\bibitem[{Moresco et~al.(2012)Moresco, Verde, Pozzetti, Jimenez and Cimatti}]{Moresco:2012by}
\bibinfo{author}{Moresco, M.}, \bibinfo{author}{Verde, L.}, \bibinfo{author}{Pozzetti, L.}, \bibinfo{author}{Jimenez, R.}, \bibinfo{author}{Cimatti, A.}, \bibinfo{year}{2012}.
\newblock \bibinfo{title}{{New constraints on cosmological parameters and neutrino properties using the expansion rate of the Universe to z{\textasciitilde}1.75}}.
\newblock \bibinfo{journal}{JCAP} \bibinfo{volume}{07}, \bibinfo{pages}{053}.
\newblock \DOIprefix\doi{10.1088/1475-7516/2012/07/053}, \href{http://arxiv.org/abs/1201.6658}{\tt arXiv:1201.6658}.
\bibitem[{Mukherjee(2016)}]{Mukherjee:2016lor}
\bibinfo{author}{Mukherjee, A.}, \bibinfo{year}{2016}.
\newblock \bibinfo{title}{{Reconstruction of interaction rate in Holographic dark energy}}.
\newblock \bibinfo{journal}{JCAP} \bibinfo{volume}{11}, \bibinfo{pages}{055}.
\newblock \DOIprefix\doi{10.1088/1475-7516/2016/11/055}, \href{http://arxiv.org/abs/1608.00400}{\tt arXiv:1608.00400}.
\bibitem[{Nastase(2016)}]{Nastase:2016sji}
\bibinfo{author}{Nastase, H.}, \bibinfo{year}{2016}.
\newblock \bibinfo{title}{{Quantum gravity and the holographic dark energy cosmology}}.
\newblock \bibinfo{journal}{JHEP} \bibinfo{volume}{04}, \bibinfo{pages}{149}.
\newblock \DOIprefix\doi{10.1007/JHEP04(2016)149}, \href{http://arxiv.org/abs/1602.03708}{\tt arXiv:1602.03708}.
\bibitem[{Nojiri and Odintsov(2006)}]{Nojiri:2005pu}
\bibinfo{author}{Nojiri, S.}, \bibinfo{author}{Odintsov, S.D.}, \bibinfo{year}{2006}.
\newblock \bibinfo{title}{{Unifying phantom inflation with late-time acceleration: Scalar phantom-non-phantom transition model and generalized holographic dark energy}}.
\newblock \bibinfo{journal}{Gen. Rel. Grav.} \bibinfo{volume}{38}, \bibinfo{pages}{1285--1304}.
\newblock \DOIprefix\doi{10.1007/s10714-006-0301-6}, \href{http://arxiv.org/abs/hep-th/0506212}{\tt arXiv:hep-th/0506212}.
\bibitem[{Nojiri and Odintsov(2011)}]{Nojiri:2010wj}
\bibinfo{author}{Nojiri, S.}, \bibinfo{author}{Odintsov, S.D.}, \bibinfo{year}{2011}.
\newblock \bibinfo{title}{{Unified cosmic history in modified gravity: from F(R) theory to Lorentz non-invariant models}}.
\newblock \bibinfo{journal}{Phys. Rept.} \bibinfo{volume}{505}, \bibinfo{pages}{59}.
\newblock \href{http://arxiv.org/abs/1011.0544}{\tt arXiv:1011.0544}.
\bibitem[{Nojiri and Odintsov(2017)}]{Nojiri:2017opc}
\bibinfo{author}{Nojiri, S.}, \bibinfo{author}{Odintsov, S.D.}, \bibinfo{year}{2017}.
\newblock \bibinfo{title}{{Covariant Generalized Holographic Dark Energy and Accelerating Universe}}.
\newblock \bibinfo{journal}{Eur. Phys. J. C} \bibinfo{volume}{77}, \bibinfo{pages}{528}.
\newblock \DOIprefix\doi{10.1140/epjc/s10052-017-5097-x}, \href{http://arxiv.org/abs/1703.06372}{\tt arXiv:1703.06372}.
\bibitem[{Nojiri et~al.(2021)Nojiri, Odintsov and Paul}]{Nojiri:2021iko}
\bibinfo{author}{Nojiri, S.}, \bibinfo{author}{Odintsov, S.D.}, \bibinfo{author}{Paul, T.}, \bibinfo{year}{2021}.
\newblock \bibinfo{title}{{Different Faces of Generalized Holographic Dark Energy}}.
\newblock \bibinfo{journal}{Symmetry} \bibinfo{volume}{13}, \bibinfo{pages}{928}.
\newblock \DOIprefix\doi{10.3390/sym13060928}, \href{http://arxiv.org/abs/2105.08438}{\tt arXiv:2105.08438}.
\bibitem[{Pavon and Zimdahl(2005)}]{Pavon:2005yx}
\bibinfo{author}{Pavon, D.}, \bibinfo{author}{Zimdahl, W.}, \bibinfo{year}{2005}.
\newblock \bibinfo{title}{{Holographic dark energy and cosmic coincidence}}.
\newblock \bibinfo{journal}{Phys. Lett. B} \bibinfo{volume}{628}, \bibinfo{pages}{206--210}.
\newblock \DOIprefix\doi{10.1016/j.physletb.2005.08.134}, \href{http://arxiv.org/abs/gr-qc/0505020}{\tt arXiv:gr-qc/0505020}.
\bibitem[{Pourhassan et~al.(2018)Pourhassan, Bonilla, Faizal and Abreu}]{Pourhassan:2017cba}
\bibinfo{author}{Pourhassan, B.}, \bibinfo{author}{Bonilla, A.}, \bibinfo{author}{Faizal, M.}, \bibinfo{author}{Abreu, E.M.C.}, \bibinfo{year}{2018}.
\newblock \bibinfo{title}{{Holographic Dark Energy from Fluid/Gravity Duality Constraint by Cosmological Observations}}.
\newblock \bibinfo{journal}{Phys. Dark Univ.} \bibinfo{volume}{20}, \bibinfo{pages}{41--48}.
\newblock \DOIprefix\doi{10.1016/j.dark.2018.02.006}, \href{http://arxiv.org/abs/1704.03281}{\tt arXiv:1704.03281}.
\bibitem[{Rudra(2023)}]{Rudra:2022qbv}
\bibinfo{author}{Rudra, P.}, \bibinfo{year}{2023}.
\newblock \bibinfo{title}{{Ricci-cubic holographic dark energy}}.
\newblock \bibinfo{journal}{Phys. Dark Univ.} \bibinfo{volume}{42}, \bibinfo{pages}{101307}.
\newblock \DOIprefix\doi{10.1016/j.dark.2023.101307}, \href{http://arxiv.org/abs/2206.03490}{\tt arXiv:2206.03490}.
\bibitem[{Saridakis(2008)}]{Saridakis:2007cy}
\bibinfo{author}{Saridakis, E.N.}, \bibinfo{year}{2008}.
\newblock \bibinfo{title}{{Restoring holographic dark energy in brane cosmology}}.
\newblock \bibinfo{journal}{Phys. Lett. B} \bibinfo{volume}{660}, \bibinfo{pages}{138--143}.
\newblock \DOIprefix\doi{10.1016/j.physletb.2008.01.004}, \href{http://arxiv.org/abs/0712.2228}{\tt arXiv:0712.2228}.
\bibitem[{Saridakis(2018)}]{Saridakis:2017rdo}
\bibinfo{author}{Saridakis, E.N.}, \bibinfo{year}{2018}.
\newblock \bibinfo{title}{{Ricci-Gauss-Bonnet holographic dark energy}}.
\newblock \bibinfo{journal}{Phys. Rev. D} \bibinfo{volume}{97}, \bibinfo{pages}{064035}.
\newblock \DOIprefix\doi{10.1103/PhysRevD.97.064035}, \href{http://arxiv.org/abs/1707.09331}{\tt arXiv:1707.09331}.
\bibitem[{Saridakis(2020)}]{Saridakis:2020zol}
\bibinfo{author}{Saridakis, E.N.}, \bibinfo{year}{2020}.
\newblock \bibinfo{title}{{Barrow holographic dark energy}}.
\newblock \bibinfo{journal}{Phys. Rev. D} \bibinfo{volume}{102}, \bibinfo{pages}{123525}.
\newblock \DOIprefix\doi{10.1103/PhysRevD.102.123525}, \href{http://arxiv.org/abs/2005.04115}{\tt arXiv:2005.04115}.
\bibitem[{Saridakis et~al.(2018)Saridakis, Bamba, Myrzakulov and Anagnostopoulos}]{Saridakis:2018unr}
\bibinfo{author}{Saridakis, E.N.}, \bibinfo{author}{Bamba, K.}, \bibinfo{author}{Myrzakulov, R.}, \bibinfo{author}{Anagnostopoulos, F.K.}, \bibinfo{year}{2018}.
\newblock \bibinfo{title}{{Holographic dark energy through Tsallis entropy}}.
\newblock \bibinfo{journal}{JCAP} \bibinfo{volume}{12}, \bibinfo{pages}{012}.
\newblock \DOIprefix\doi{10.1088/1475-7516/2018/12/012}, \href{http://arxiv.org/abs/1806.01301}{\tt arXiv:1806.01301}.
\bibitem[{Saridakis et~al.(2021)}]{CANTATA:2021asi}
\bibinfo{author}{Saridakis, E.N.}, et~al. (\bibinfo{collaboration}{CANTATA}), \bibinfo{year}{2021}.
\newblock \bibinfo{title}{{Modified Gravity and Cosmology. An Update by the CANTATA Network}}.
\newblock \bibinfo{publisher}{Springer}.
\newblock \DOIprefix\doi{10.1007/978-3-030-83715-0}, \href{http://arxiv.org/abs/2105.12582}{\tt arXiv:2105.12582}.
\bibitem[{Schwarz(1978)}]{Schwarz:1978tpv}
\bibinfo{author}{Schwarz, G.}, \bibinfo{year}{1978}.
\newblock \bibinfo{title}{{Estimating the Dimension of a Model}}.
\newblock \bibinfo{journal}{Annals Statist.} \bibinfo{volume}{6}, \bibinfo{pages}{461--464}.
\bibitem[{Scolnic et~al.(2022)}]{Scolnic:2021amr}
\bibinfo{author}{Scolnic, D.}, et~al., \bibinfo{year}{2022}.
\newblock \bibinfo{title}{{The Pantheon+ Analysis: The Full Data Set and Light-curve Release}}.
\newblock \bibinfo{journal}{Astrophys. J.} \bibinfo{volume}{938}, \bibinfo{pages}{113}.
\newblock \DOIprefix\doi{10.3847/1538-4357/ac8b7a}, \href{http://arxiv.org/abs/2112.03863}{\tt arXiv:2112.03863}.
\bibitem[{Setare(2006)}]{Setare:2006wh}
\bibinfo{author}{Setare, M.R.}, \bibinfo{year}{2006}.
\newblock \bibinfo{title}{{Interacting holographic dark energy model in non-flat universe}}.
\newblock \bibinfo{journal}{Phys. Lett. B} \bibinfo{volume}{642}, \bibinfo{pages}{1--4}.
\newblock \DOIprefix\doi{10.1016/j.physletb.2006.09.027}, \href{http://arxiv.org/abs/hep-th/0609069}{\tt arXiv:hep-th/0609069}.
\bibitem[{Setare and Saridakis(2008)}]{Setare:2008hm}
\bibinfo{author}{Setare, M.R.}, \bibinfo{author}{Saridakis, E.N.}, \bibinfo{year}{2008}.
\newblock \bibinfo{title}{{Correspondence between Holographic and Gauss-Bonnet dark energy models}}.
\newblock \bibinfo{journal}{Phys. Lett. B} \bibinfo{volume}{670}, \bibinfo{pages}{1--4}.
\newblock \DOIprefix\doi{10.1016/j.physletb.2008.10.029}, \href{http://arxiv.org/abs/0810.3296}{\tt arXiv:0810.3296}.
\bibitem[{Setare and Vagenas(2008)}]{Setare:2008bb}
\bibinfo{author}{Setare, M.R.}, \bibinfo{author}{Vagenas, E.C.}, \bibinfo{year}{2008}.
\newblock \bibinfo{title}{{Thermodynamical Interpretation of the Interacting Holographic Dark Energy Model in a non-flat Universe}}.
\newblock \bibinfo{journal}{Phys. Lett. B} \bibinfo{volume}{666}, \bibinfo{pages}{111--115}.
\newblock \DOIprefix\doi{10.1016/j.physletb.2008.07.013}, \href{http://arxiv.org/abs/0801.4478}{\tt arXiv:0801.4478}.
\bibitem[{Shannon(1948)}]{shannon1948claude}
\bibinfo{author}{Shannon, C.E.}, \bibinfo{year}{1948}.
\newblock \bibinfo{title}{Claude elwood shannon}.
\newblock \bibinfo{journal}{Bell Syst. Tech. J} \bibinfo{volume}{27}, \bibinfo{pages}{379--423}.
\bibitem[{Shekh(2021)}]{Shekh:2021ule}
\bibinfo{author}{Shekh, S.H.}, \bibinfo{year}{2021}.
\newblock \bibinfo{title}{{Models of holographic dark energy in $f(Q)$ gravity}}.
\newblock \bibinfo{journal}{Phys. Dark Univ.} \bibinfo{volume}{33}, \bibinfo{pages}{100850}.
\newblock \DOIprefix\doi{10.1016/j.dark.2021.100850}.
\bibitem[{Sheykhi(2009)}]{Sheykhi:2009dz}
\bibinfo{author}{Sheykhi, A.}, \bibinfo{year}{2009}.
\newblock \bibinfo{title}{{Interacting holographic dark energy in Brans-Dicke theory}}.
\newblock \bibinfo{journal}{Phys. Lett. B} \bibinfo{volume}{681}, \bibinfo{pages}{205--209}.
\newblock \DOIprefix\doi{10.1016/j.physletb.2009.10.011}, \href{http://arxiv.org/abs/0907.5458}{\tt arXiv:0907.5458}.
\bibitem[{Susskind(1995)}]{Susskind:1994vu}
\bibinfo{author}{Susskind, L.}, \bibinfo{year}{1995}.
\newblock \bibinfo{title}{{The World as a hologram}}.
\newblock \bibinfo{journal}{J. Math. Phys.} \bibinfo{volume}{36}, \bibinfo{pages}{6377--6396}.
\newblock \DOIprefix\doi{10.1063/1.531249}, \href{http://arxiv.org/abs/hep-th/9409089}{\tt arXiv:hep-th/9409089}.
\bibitem[{Suwa and Nihei(2010)}]{Suwa:2009gm}
\bibinfo{author}{Suwa, M.}, \bibinfo{author}{Nihei, T.}, \bibinfo{year}{2010}.
\newblock \bibinfo{title}{{Observational constraints on the interacting Ricci dark energy model}}.
\newblock \bibinfo{journal}{Phys. Rev. D} \bibinfo{volume}{81}, \bibinfo{pages}{023519}.
\newblock \DOIprefix\doi{10.1103/PhysRevD.81.023519}, \href{http://arxiv.org/abs/0911.4810}{\tt arXiv:0911.4810}.
\bibitem[{{Torrado} and {Lewis}(2019)}]{2019ascl.soft10019T}
\bibinfo{author}{{Torrado}, J.}, \bibinfo{author}{{Lewis}, A.}, \bibinfo{year}{2019}.
\newblock \bibinfo{title}{{Cobaya: Bayesian analysis in cosmology}}.
\newblock \bibinfo{howpublished}{Astrophysics Source Code Library, record ascl:1910.019}.
\newblock \href{http://arxiv.org/abs/1910.019}{\tt arXiv:1910.019}.
\bibitem[{Torrado and Lewis(2021)}]{Torrado:2020dgo}
\bibinfo{author}{Torrado, J.}, \bibinfo{author}{Lewis, A.}, \bibinfo{year}{2021}.
\newblock \bibinfo{title}{{Cobaya: Code for Bayesian Analysis of hierarchical physical models}}.
\newblock \bibinfo{journal}{JCAP} \bibinfo{volume}{05}, \bibinfo{pages}{057}.
\newblock \DOIprefix\doi{10.1088/1475-7516/2021/05/057}, \href{http://arxiv.org/abs/2005.05290}{\tt arXiv:2005.05290}.
\bibitem[{Tsallis and Cirto(2013)}]{Tsallis:2013}
\bibinfo{author}{Tsallis, C.}, \bibinfo{author}{Cirto, L.J.L.}, \bibinfo{year}{2013}.
\newblock \bibinfo{title}{{Black hole thermodynamical entropy}}.
\newblock \bibinfo{journal}{Eur. Phys. J. C} \bibinfo{volume}{73}.
\newblock \DOIprefix\doi{10.1140/epjc/s10052-013-2487-6}.
\bibitem[{Wang et~al.(2005)Wang, Gong and Abdalla}]{Wang:2005jx}
\bibinfo{author}{Wang, B.}, \bibinfo{author}{Gong, Y.g.}, \bibinfo{author}{Abdalla, E.}, \bibinfo{year}{2005}.
\newblock \bibinfo{title}{{Transition of the dark energy equation of state in an interacting holographic dark energy model}}.
\newblock \bibinfo{journal}{Phys. Lett. B} \bibinfo{volume}{624}, \bibinfo{pages}{141--146}.
\newblock \DOIprefix\doi{10.1016/j.physletb.2005.08.008}, \href{http://arxiv.org/abs/hep-th/0506069}{\tt arXiv:hep-th/0506069}.
\bibitem[{Wang et~al.(2017)Wang, Wang and Li}]{Wang:2016och}
\bibinfo{author}{Wang, S.}, \bibinfo{author}{Wang, Y.}, \bibinfo{author}{Li, M.}, \bibinfo{year}{2017}.
\newblock \bibinfo{title}{{Holographic Dark Energy}}.
\newblock \bibinfo{journal}{Phys. Rept.} \bibinfo{volume}{696}, \bibinfo{pages}{1--57}.
\newblock \DOIprefix\doi{10.1016/j.physrep.2017.06.003}, \href{http://arxiv.org/abs/1612.00345}{\tt arXiv:1612.00345}.
\bibitem[{Yarahmadi and Salehi(2024)}]{Yarahmadi:2024oqv}
\bibinfo{author}{Yarahmadi, M.}, \bibinfo{author}{Salehi, A.}, \bibinfo{year}{2024}.
\newblock \bibinfo{title}{{Alleviating the Hubble tension using the Barrow holographic dark energy cosmology with Granda{\textendash}Oliveros IR cut-off}}.
\newblock \bibinfo{journal}{Mon. Not. Roy. Astron. Soc.} \bibinfo{volume}{534}, \bibinfo{pages}{3055--3067}.
\newblock \DOIprefix\doi{10.1093/mnras/stae2257}.
\bibitem[{Zhai et~al.(2011)Zhai, Zhang and Liu}]{Zhai:2011pp}
\bibinfo{author}{Zhai, Z.X.}, \bibinfo{author}{Zhang, T.J.}, \bibinfo{author}{Liu, W.B.}, \bibinfo{year}{2011}.
\newblock \bibinfo{title}{{Constraints on $\Lambda(t)$CDM models as holographic and agegraphic dark energy with the observational Hubble parameter data}}.
\newblock \bibinfo{journal}{JCAP} \bibinfo{volume}{08}, \bibinfo{pages}{019}.
\newblock \DOIprefix\doi{10.1088/1475-7516/2011/08/019}, \href{http://arxiv.org/abs/1109.1661}{\tt arXiv:1109.1661}.
\bibitem[{Zhang et~al.(2015)Zhang, Zhao, Li and Zhang}]{Zhang:2015rha}
\bibinfo{author}{Zhang, J.F.}, \bibinfo{author}{Zhao, M.M.}, \bibinfo{author}{Li, Y.H.}, \bibinfo{author}{Zhang, X.}, \bibinfo{year}{2015}.
\newblock \bibinfo{title}{{Neutrinos in the holographic dark energy model: constraints from latest measurements of expansion history and growth of structure}}.
\newblock \bibinfo{journal}{JCAP} \bibinfo{volume}{04}, \bibinfo{pages}{038}.
\newblock \DOIprefix\doi{10.1088/1475-7516/2015/04/038}, \href{http://arxiv.org/abs/1502.04028}{\tt arXiv:1502.04028}.
\bibitem[{Zhang(2009)}]{Zhang:2009un}
\bibinfo{author}{Zhang, X.}, \bibinfo{year}{2009}.
\newblock \bibinfo{title}{{Holographic Ricci dark energy: Current observational constraints, quintom feature, and the reconstruction of scalar-field dark energy}}.
\newblock \bibinfo{journal}{Phys. Rev. D} \bibinfo{volume}{79}, \bibinfo{pages}{103509}.
\newblock \DOIprefix\doi{10.1103/PhysRevD.79.103509}, \href{http://arxiv.org/abs/0901.2262}{\tt arXiv:0901.2262}.
\bibitem[{Zhang et~al.(2012)Zhang, Li, Li, Zhang and Li}]{Zhang:2012uu}
\bibinfo{author}{Zhang, Z.}, \bibinfo{author}{Li, S.}, \bibinfo{author}{Li, X.D.}, \bibinfo{author}{Zhang, X.}, \bibinfo{author}{Li, M.}, \bibinfo{year}{2012}.
\newblock \bibinfo{title}{{Revisit of the Interaction between Holographic Dark Energy and Dark Matter}}.
\newblock \bibinfo{journal}{JCAP} \bibinfo{volume}{06}, \bibinfo{pages}{009}.
\newblock \DOIprefix\doi{10.1088/1475-7516/2012/06/009}, \href{http://arxiv.org/abs/1204.6135}{\tt arXiv:1204.6135}.
\bibitem[{Zhao et~al.(2017)Zhao, He, Zhang and Zhang}]{Zhao:2017urm}
\bibinfo{author}{Zhao, M.M.}, \bibinfo{author}{He, D.Z.}, \bibinfo{author}{Zhang, J.F.}, \bibinfo{author}{Zhang, X.}, \bibinfo{year}{2017}.
\newblock \bibinfo{title}{{Search for sterile neutrinos in holographic dark energy cosmology: Reconciling Planck observation with the local measurement of the Hubble constant}}.
\newblock \bibinfo{journal}{Phys. Rev. D} \bibinfo{volume}{96}, \bibinfo{pages}{043520}.
\newblock \DOIprefix\doi{10.1103/PhysRevD.96.043520}, \href{http://arxiv.org/abs/1703.08456}{\tt arXiv:1703.08456}.

\end{thebibliography}



\end{document}